\documentclass[prd,aps,twocolumn,floats,floatfix,nofootinbib] {revtex4-1}

\usepackage{graphicx}
\usepackage{dcolumn}
\usepackage{bm}
\usepackage{graphics}
\usepackage{slashed}
\usepackage{amssymb}
\usepackage{natbib}
\usepackage{amsmath}
\usepackage{url}
\usepackage[usenames,dvipsnames,svgnames,table]{xcolor}
\usepackage{color}
\usepackage{xcolor}
\usepackage{tabularx}
\usepackage{hyperref}
\usepackage{enumitem}
\usepackage{soul}

\begin{document}
\title{Gravitational waves and kicks from the merger of\\ unequal mass, highly compact boson stars}
	
\author{
		Miguel Bezares$^{1,2}$,
		Mateja Bo\v skovi\'c$^{1,2}$,
		Steven Liebling$^{3}$,
		Carlos Palenzuela$^{4}$,
		Paolo  Pani$^{5}$,
 	    Enrico Barausse$^{1,2}$
}
\affiliation{${^1}$SISSA, Via Bonomea 265, 34136 Trieste, Italy and INFN Sezione di Trieste}
\affiliation{${^2}$IFPU - Institute for Fundamental Physics of the Universe, Via Beirut 2, 34014 Trieste, Italy}
\affiliation{${^3}$Long Island University, Brookville, New York 11548, USA}	
\affiliation{${^4}$Departament  de  F\'{\i}sica $\&$ IAC3,  Universitat  de  les  Illes  Balears  and  Institut  d'Estudis
Espacials  de  Catalunya,  Palma  de  Mallorca,  Baleares  E-07122,  Spain}
\affiliation{$^{5}$Dipartimento di Fisica, ``Sapienza'' Universit\`a di Roma \& Sezione INFN Roma1, P.A. Moro 5, 00185, Roma, Italy}

\begin{abstract}
Boson stars have attracted much attention in recent decades as simple, self-consistent models of compact objects and also as self-gravitating structures formed in some dark-matter scenarios.
Direct detection of these hypothetical objects through electromagnetic signatures would be unlikely because their bosonic constituents are not expected to interact significantly with ordinary matter and radiation.
However, binary boson stars might form and coalesce emitting a detectable gravitational wave signal which might distinguish them from ordinary compact object binaries containing black holes and neutron stars.
We study the merger of two boson stars by numerically evolving the fully relativistic Einstein-Klein-Gordon equations for a complex scalar field with a solitonic potential that generates very compact boson stars.
Owing to the steep mass-radius diagram, we can study the dynamics and gravitational radiation from unequal-mass binary boson stars with mass ratios up to $q\approx23$ without the difficulties encountered when evolving binary black holes with large mass ratios. 
Similar to the previously-studied equal-mass case, our numerical evolutions of the merger produce either a nonspinning boson star or a spinning black hole, depending on the initial masses and on the binary angular momentum. We do not find any evidence of synchronized scalar clouds forming around either the remnant spinning black hole or around the remnant boson stars. Interestingly,  in contrast to the equal-mass case, one of the mechanisms to dissipate  angular momentum is now asymmetric, and leads to large kick velocities (up to a few $10^4\,{\rm km/s}$) which could produce wandering remnant boson stars. We also compare the gravitational wave signals predicted from boson star binaries with those from black hole binaries, and comment on the detectability of the differences with ground interferometers.
\end{abstract}

\maketitle
\tableofcontents

\section{Introduction}
We are well into the era of gravitational wave~(GW) astronomy with
the rapidly growing catalog of GW events detected by the LIGO-Virgo collaboration~\cite{LIGOScientific:2014pky,TheVirgo:2014hva}.

With the very recent release of the third GW transient catalog~\cite{LIGOScientific:2021djp}, the total number of reported coalescences increased to $90$. 
Some of the more remarkable events detected to date include:
\begin{itemize}[noitemsep,topsep=0pt]
  \item  GW190412~\cite{LIGOScientific:2020stg}, a binary black hole~(BBH) with asymmetric component masses, showing evidence for higher harmonics in its GW signal;
  \item  GW190425~\cite{LIGOScientific:2020aai}, identified with a binary neutron star~(NS) merger lacking evidence of an electromagnetic counterpart;
  \item  GW190521~\cite{LIGOScientific:2020iuh}, a BBH with a total mass greater than 150 solar masses, which is the most massive binary yet detected, 
         in which the posterior distribution of the primary mass is nearly entirely in the pair-instability supernova mass gap where BHs are not expected to form from the collapse of massive stars;
  \item  GW190814~\cite{LIGOScientific:2020zkf}, a highly asymmetric system  consistent with the merger of a 23 solar mass black hole~(BH) with a 2.6 solar mass compact object, making the latter either the lightest BH or the heaviest NS observed in a compact binary; 
  \item  GW200105 and GW200115~\cite{LIGOScientific:2021qlt}, which are the first detections consistent with a NS-BH merger. 
   \end{itemize}
The planned upgrades by the LIGO-Virgo collaboration and the addition of the KAGRA detector~\cite{KAGRA:2020agh} promise even more exciting observations in the future.

A primary target of GW observations is the merger of very compact objects, with BHs and NSs being the most natural candidates. However, a number of other hypothetical compact objects have been proposed, called exotic compact objects~(ECOs)~\cite{Giudice:2016zpa,Cardoso:2019rvt}. The motivations for various ECOs arise both in beyond-Standard-Model theories and in modified-gravity scenarios, and some of the most popular models include fuzzballs~\cite{Mathur:2005zp}, gravastars~\cite{Mazur:2001fv}, wormholes~\cite{Damour:2007ap}, anisotropic stars~\cite{Raposo:2018rjn}, and boson stars~(BSs)~\cite{Ruffini:1969qy}. Phenomenological studies of ECOs are required to perform actual searches for their signatures. 
No evidence for such ECOs has yet been found, but, because they are expected to be too dim electromagnetically, it is mostly through GW detections that we can hope to observe them~\cite{Cardoso:2019rvt}.

In this work we study BSs, which are solutions of the Einstein equations coupled to a complex scalar field with a harmonic time dependence  describing a macroscopic wave-function of a Bose-Einstein condensate (see Ref.~\cite{Schunck:2003kk,liebpa,Visinelli:2021uve} for reviews). BSs are particularly promising as possible astrophysical objects because:
(i)~a formation mechanism for BSs has been identified, known as gravitational cooling~\cite{Seidel:1993zk,Guzman:2006yc}, whereby BSs can be produced from arbitrary scalar field configurations,
(ii)~their stability properties resemble those of NSs so that static BSs below a critical mass are radially stable~\cite{Colpi:1986ye,liebpa,mace},
 and finally 
(iii)~BSs have been invoked in open problems in cosmological and particle physics, such as the nature of the dark matter  and the possibility of early Universe remnants. 
For instance, the idea that dark matter is composed of ultra-light bosonic fields has received significant attention recently~\cite{Hui:2016ltb,Marsh:2016rep,Hui:2021tkt}. Although leading candidates for this kind of dark matter are real scalars that are organized in time-dependent configurations~\cite{Seidel:1991zh}, BSs can serve as a proxy for such configurations~\cite{Brito:2015yfh}. Some of these scenarios can allow for compact BSs (or similar objects) to be produced in the early Universe~\cite{Troitsky:2015mda,Krippendorf:2018tei,Levkov:2018kau,Cotner:2019ykd,Amin:2019ums,Widdicombe:2018oeo,Arvanitaki:2019rax}

Collisions of BSs have been studied extensively, including:  head-on and orbital mergers of mini-BSs~\cite{pale1,pale2}, 
head-on mergers of oscillatons~\cite{Brito:2015yfh,Helfer:2018vtq},  orbital collisions of solitonic BSs~\cite{bezpalen,PhysRevD.96.104058},
and head-on and orbital mergers of Proca stars~\cite{Brito:2015pxa,Sanchis-Gual:2017bhw} as a possible alternative explanation of the GW190521 event~\cite{PhysRevD.99.024017,Bustillo:2020syj}. 
The merger of ECOs can be studied within various dark matter scenarios as well, as for example: 
      mergers between a NSs and a star made of axions, one of the most popular dark matter type candidates~\cite{Dietrich:2018bvi,Dietrich:2018jov,Clough:2018exo}, 
      mergers of dark stars composed of bosonic fields~\cite{Bezares:2018qwa}, 
   or mergers of binary NSs containing a small fraction of dark matter~\cite{Bezares:2019jcb} modeled using fermion-BSs~\cite{suspalen}.

Motivated by the recent GW detections of very unequal mass binary mergers, we study here the coalescence of unequal mass BS binaries, focusing on their dynamics and GW radiation.
As in our previous works~\cite{bezpalen,PhysRevD.96.104058,Bezares:2018qwa}, we adopt the nontopological solitonic BS potential~\cite{frie} to construct our asymmetric binaries because: 
(i)~it allows for very compact configurations that reach a maximum compactness (see below for its definition) in the stable branch of approximately $C\approx 0.35$~\cite{Boskovic:2021nfs, Cardoso:2021ehg},
and (ii)~one can construct binaries with a large mass ratio. Indeed, defining the mass ratio $q\equiv m_{1}/m_{2}$ such that $m_{1}>m_{2},$ we can produce compact binaries with a mass ratio ranging\footnote{Solitonic BSs in general admit two stable and two unstable branches~\cite{Tamaki:2011zza,Boskovic:2021nfs}. Here we focus on the more massive stable branch, while the other stable branch corresponds to the weak-field regime of mini BSs for our choice of the potential parameters~\cite{Boskovic:2021nfs}.} approximately from $1$ to $45$.  Here, we focus on binaries within the range $q \in [2,23]$. We note that in contrast to the difficulties encountered when evolving BBH with large 
mass ratios~\cite{Gonzalez:2008bi,Lousto:2010qx,Lousto:2010ut,BBHune}, these evolutions require no change to the choice of coordinates, namely gamma-driver shift condition, nor an exceptionally high resolution. The reason for this difference is because the radii of solitonic BSs even with vastly different masses are of the same order, whereas the radius of the BH scales linearly with the mass, and therefore a large mass ratio in a BH binary necessarily implies a large separation of length scales.

Our mergers of unequal mass solitonic BSs produce either a non-rotating BS or a spinning BH, as in the equal-mass cases~\cite{PhysRevD.96.104058}. In the former cases, all the angular momentum is emitted to infinity through scalar field and GW radiation, while in the latter case, after performing a very long-term simulation, we find no indication of a scalar cloud synchronized with the rotation of the remnant BH, as found in Ref.~\cite{Sanchis-Gual:2020mzb}.
For one of our simulations with large angular momentum, a blob of scalar field is ejected after the merger, producing a significant kick velocity of the remnant. Note that, this blob ejection has already been observed in solitonic BS binaries of equal mass~\cite{PhysRevD.96.104058}. Additionally, we study the dynamics and GW radiation of a binary composed of a BS and an anti-boson~(aBS) star, i.e.  with the opposite frequency, allowing some annihilation of the Noether charge during the merger. 

This work is organized as follows: in Sec.~\ref{stp}, we review the evolution equations describing BSs, followed by  the construction of initial data for binary BSs and numerical implementation. In Sec.~\ref{dynamics}, the coalescence of unequal-mass BS binaries is studied in detail. The GWs produced by these systems are explored in Sec.~\ref{gwsect}, in particular, analyzing the imprint of higher-order modes in the signal and the post-merger frequencies of the remnant's signal. In Sec.~\ref{conclu}, we summarize our results. 
We use geometric units in which $G=1$ and $c=1,$ unless otherwise stated.
	
\section{Setup}\label{stp}
In this section, we briefly summarize the evolution equations describing a self-gravitating (complex) scalar field  and the construction of binary BSs in quasicircular orbits that constitute the initial data.
We also outline the numerical methods and grid setup employed to perform the simulations. Notice that our setup is very similar to the one used in	Ref.~\cite{PhysRevD.96.104058} (hereafter Paper~I) for studying equal-mass binary BSs.
	
\subsection{Einstein-Klein-Gordon equations}

Self-gravitating (complex) scalar-fields are described by the Einstein-Klein-Gordon~(EKG) equations
\begin{eqnarray}
R_{ab} - \frac{1}{2}g_{ab} R &=& 
8\pi \, T_{ab}\,,\\
g^{ab} \nabla_a \nabla_b \Phi &=& \frac{dV}{d |\Phi|^2} \Phi\,,
\label{EKG_equations}
\end{eqnarray}
where $R_{ab}$ is the Ricci tensor associated with the metric $g_{ab}$, $\Phi$ is a minimally coupled, complex scalar field, and $V \left(|\Phi|^2\right)$ is its associated self-interaction potential. The stress-energy tensor $T_{ab}$ for
the complex scalar field is given by
\begin{equation}
T_{ab} = \nabla_a \Phi \nabla_b \Phi^* +
\nabla_a \Phi^* \nabla_b \Phi -
g_{ab} \left[ \nabla^c \Phi \nabla_c \Phi^*
+ V\left(|\Phi|^2\right) \right],  \nonumber
\end{equation}
where $\Phi^*$ is the complex conjugate of $\Phi$. Different BS models are classified according to their scalar self-potential $V\left(|\Phi|^2\right)$. Here we focus on the solitonic potential~\cite{frie}, which allows for highly compact BSs and is given by
\begin{equation}
V\left(|\Phi|^2\right)=m_b^2|\Phi|^2\left[1-\frac{2|\Phi|^2}{\sigma_0^2}\right]^2\,,\label{potential}
\end{equation}
where $m_b$ and $\sigma_0$ are two free parameters. In our units (in which the scalar field is dimensionless), $m_b$ has the dimensions of an inverse length, $m_b\hbar$ is the bare mass of the scalar field, whereas $\sigma_0$ is dimensionless. 
We define
$\lambda \equiv \sigma_0 \sqrt{8 \pi}$ and set $m_b\lambda=1$ for the rest of the paper.
However, in some occurrences we shall re-insert the proper factors of $m_b \lambda$.

In the complex-$\Phi$ space the potential has the typical Mexican-hat shape, with a maximum at $\Phi=0$ and a minimum (degenerate vacuum) at $|\Phi|=\sigma_0/\sqrt{2}$. When $\sigma_0\ll1$ the scalar profile is roughly constant within the star and steeply vanishes over a lengthscale $\sim 1/m_b$~\cite{Kesden:2004qx,Boskovic:2021nfs}.

Due to the $U(1)$ invariance of the EKG action, BSs admit a conserved Noether charge current
\begin{equation}
j^{a} = i g^{ab} (\Phi^*\,\nabla_{b}\Phi - \Phi\,\nabla_{b} \Phi^*)\,.
\end{equation}
The spatial integral of the time component of this current defines the conserved Noether charge, $N$,
which can be interpreted as the number of bosonic particles in the star~\cite{liebpa}.

\subsection{Binary initial data}\label{subsec:IDBS}

\begin{figure}
\centering
\includegraphics[width=0.45\textwidth]{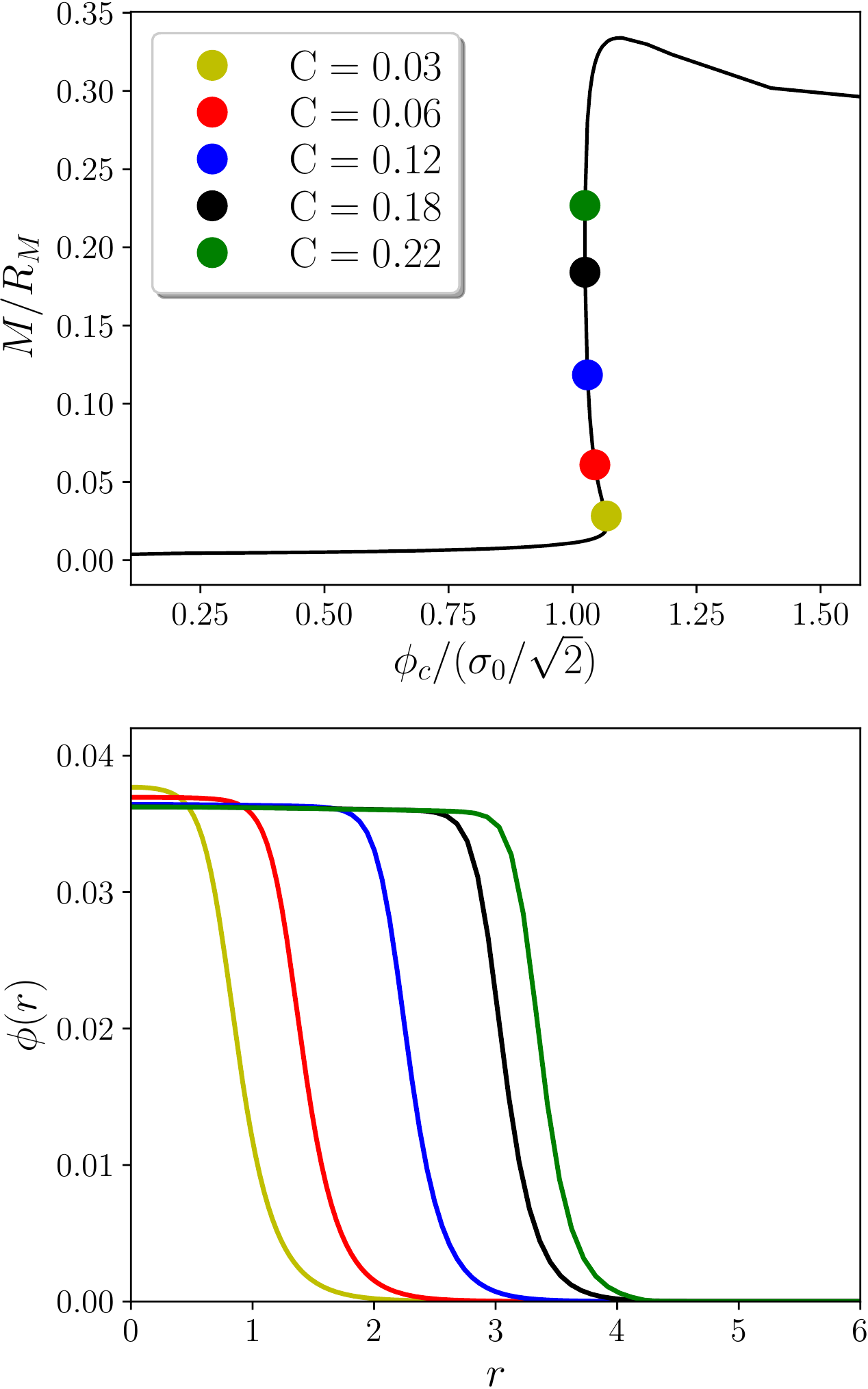}
\caption{ {\em  Isolated BS solutions}.  (\textbf{Top}) Compactness as a function of the central value of the scalar field $\phi_c$ for isolated, nonrotating BSs with $\sigma_0=0.05$. Circular markers refer to the equilibrium configurations used in this work to construct initial data for BS binaries [cf.\ Table~\ref{isolatedbs}]. The radius $R_M$ is defined as that containing $99\%$ of the mass of the star. (\textbf{Bottom}) Profile of the scalar field as a function of the isotropic radius for the different configurations.}
\label{fig:isolatedBS}
\end{figure}

The procedure to construct the initial data for a binary BS is the same as in Paper~I, that is, a superposition of two boosted, isolated, solitonic BSs. 

The solution of a single solitonic BS is constructed as described in Ref.~\cite{mace}, by adopting the usual harmonic ansatz for the scalar field $\Phi = \phi(r)\,e^{-i\omega t}$ with a real frequency $\omega$. Assuming stationarity and spherical symmetry, the EKG equations reduce to a set of ordinary differential equations which can be solved numerically with a shooting method.  Integrating from the center with a given central value  of the scalar field $\phi_c$ and frequency $\omega$, one looks
for solutions satisfying regularity and boundary conditions.
The resulting BS equilibrium configurations can be characterized by their mass and radius. 
However, because the scalar field only vanishes asymptotically as it decays exponentially, the definition of its radius is necessarily somewhat ambiguous.
Following previous work, we can define the effective radius $R_M$ as the radius within which $99\%$ of the total mass is contained, i.e.\ $m(R_M)=0.99M$. Consequently, we define the compactness as $C\equiv M/R_M$. As a reference, the compactness for a Schwarzschild BH is $C=0.5$ and $C\approx 0.1-0.2$ for NSs. 
In numerical simulations, it is however more convenient to estimate the radius of the final remnant through the radius that contains $99\%$ of the Noether charge, $R_N$, so we will use this definition when required. The radius of the remnant is calculated with respect to its center of mass.

The maximum mass of static configurations in this model is 
\begin{equation}
 M_{\rm max} \approx 5 M_\odot \left[\frac{10^{-12}}{\sigma_0}\right]^2 \left[\frac{500\,{\rm GeV}}{m_b\hbar}\right] \,,\label{Mmax}
\end{equation}
where the scaling with $m_b^{-1}$ is exact, whereas the scaling with $\sigma_0^{-2}$ is approximately valid only in the $\sigma_0\ll1$ limit. Thus, depending on $(m_b,\sigma_0)$ the model supports self-gravitating configurations across a wide mass range.

Paper~I presented a sequence of isolated BS solutions characterized by the central value of the scalar field $\phi_c$ that we use to construct our unequal mass binaries here. In the top panel of Fig.~\ref{fig:isolatedBS}, the compactness $C$ is shown as a function of $\phi_c$.
The circular markers denote the five representative BSs employed in this paper. The bottom panel of Fig.~\ref{fig:isolatedBS} displays the radial profile of the scalar field for these isolated solutions, while Table~\ref{isolatedbs} lists the key properties of these configurations.
	
Notice that these solutions can be rewritten in terms of the following dimensionless quantities~\cite{mace}
\begin{eqnarray}
M (m_b \lambda) ,~~  N (m_b \lambda)^2 ,~~ r (m_b \lambda) ,~~ \omega/(m_b \lambda) \,,
\label{eq:scaling}
\end{eqnarray}
recalling 
that $\lambda = \sigma_0 \sqrt{8 \pi}$. In terms of these parameters, the equations become independent of $m_b$, and hence $m_b$ serves to set the units of the physical solution. Again, the linear scaling in $m_b$ in the above expressions is exact, whereas that with respect to $\lambda$ is approximately valid only in the $\sigma_0\ll1$ limit. For the chosen value, $\sigma_0=0.05$, this scaling is already a good approximation, and so smaller values of $\sigma_0$ can be studied simply by applying such a rescaling.
Here we restrict ourselves to $\sigma_0=0.05$, which sufficiently fulfills the condition $\sigma_0\ll1$ and also allows for very compact, stable configurations. 
\begin{table*}
		\begin{ruledtabular}
			\begin{tabular}{c||cccccc||cc}
				$C$ & $\phi_c/(\sigma_0/\sqrt{2})$ & $Mm_b \lambda$ & $N(m_b \lambda)^2$ & $(R_M,R_N)m_b  \lambda$ & $\omega/(m_b \lambda)$  & & $I/M^3$ & $k_{\rm tidal}$ 
				\\ \hline\hline 
				0.03 & 1.065     & 0.0463   &  0.01653   & (1.507,\,1.380)       & 2.129620346  & & $245.3$ & $136494$ \\ 
				0.06 & 1.045 & 0.1238   & 0.0605   & (2.0334,\,1.8288) & 1.545745909  & & $84.9$ & $8420$ \\ 
				0.12 & 1.030 & 0.3650   & 0.2551   & (3.0831,\,2.8360) & 1.066612350  & & $27.8$  & $332$ \\ 
				0.18 & 1.025 & 0.7835   & 0.7193   & (4.2572,\,3.9960) & 0.790449025  & & $12.5$  & $41$  \\ 
				0.22 & 1.025 & 1.0736   & 1.1147   & (4.9647,\,4.7068) & 0.685760351  & & $8.34$  & $20$  \\
			\end{tabular}
			\caption{{\em Characteristics of solitonic BS models with $\sigma_0=0.05$.} The table shows: 
     $\bullet$~the compactness $C$, 
     $\bullet$~the central value of the scalar field $\phi_c/(\sigma_0/\sqrt{2})$, 
     $\bullet$~the ADM mass $Mm_b\lambda$, 
     $\bullet$~the Noether charge $N\left(m_b\lambda\right)^2$, 
     $\bullet$~the radius of the star (the radius containing $99\%$ of either the mass or of the Noether charge for $R_M$ or $R_N$, respectively), 
 and $\bullet$~the angular frequency of the field in the complex plane, $\omega/\left(m_b\lambda\right)$, in dimensionless units. 
In the last two columns, we give 
     $\bullet$~the normalized, Newtonian, moment of inertia (where $I=\int \rho^2\, dm$, where $\rho$ is the distance from the axis of rotation), 
and $\bullet$~the dimensionless tidal Love number, $k_{\rm tidal}$, as computed in Refs.~\cite{Cardoso:2017cfl,Sennett:2017etc}. 
For a NS with an ordinary equation of state and $C\sim 0.1$, $k_{\rm tidal}={\cal O}(200)$ while $k_{\rm tidal}=0$ for a BH. 
}\label{isolatedbs}
		\end{ruledtabular}
	\end{table*}

The initial data for the BS binary follows the procedure described in Ref.~\cite{bezpalen} and Paper~I.
Once the isolated BSs are constructed in spherical coordinates, the solution is extended to Cartesian coordinates, with the centers of the stars located at along the $y$-axis at $(0,y^j_c,0)$, so that the center of mass of the system is located at the origin.\footnote{Here we define the center of mass using the masses of isolated configurations listed in Table~\ref{isolatedbs}. Constraint violation transient will change these masses, see the discussion of ``effective" configurations below.} A Lorentz transformation is performed to boost each star along the $\pm x$-directions, and finally the boosted solutions for both  stars are superposed to obtain our binary initial data. Obviously, this superposition is only an approximate solution that does not satisfy exactly the constraints at the initial time (see Ref.~\cite{Helfer:2021brt} for a partial solution in case of  equal mass binaries of BSs). However, our evolution scheme enforces an exponential decay of this constraint violation dynamically (e.g., see Fig.~10 in Ref.~\cite{bezpalen}).
	
In contrast with Paper~I where the positions and initial velocities of each binary were anti-symmetric (i.e., velocities with the same magnitude but opposite direction), for these unequal cases we have set those parameters as follows: given an initial separation we have calculated the 2nd order post-Newtonian orbital velocity~\cite{Mirshekari:2013vb} such that the system would be in quasicircular orbit and the velocity of the center of mass would be close to zero. Then, we modify these velocities by adding a tiny amount of linear drift velocity to account for the finite initial orbital distance and higher-order relativistic effects, and fix this drift velocity such that the velocity of the binary center of mass is close to zero. The positions and velocities of each binary system considered in this work, together with other parameters of our simulations, are presented in Table~\ref{table2}.
	\begin{table*}
		\begin{ruledtabular}
			\begin{tabular}{c||cc||cccccc|cc|cccc}
				Binaries & $q$ & $\nu$ & $y_c^{(1)}$ & $y_c^{(2)}$ & $v_x^{(1)}$ & $v_x^{(2)}$ &  $M_0m_b \lambda$ & $J_0(m_b \lambda)^{2}$  & 	 $t_{\mathrm{c}}$ & $t^{\mathrm{ret}}_{\mathrm{m}}$& {\rm remnant} & $M_{\rm r}m_{b}\lambda$  &  $R_{\rm N}m_b  \lambda$ & $M_r \omega^0_{\rm r}$\\
				\hline\hline 
				C003 - C022A& 23.2 & 0.039 & $-$9.58 & 0.42 & $-$0.34  & 0.02   &  1.16 & 0.229  & 790 & 811 & BS & 1.07 & 4.50 &  0.218 \\
				C003 - C022 & 23.2 & 0.039 & $-$9.58 & 0.42 & $-$0.34  & 0.02   &  1.16 & 0.229  & 790 & 808 & BS & 1.13 & 4.76 &  0.228 \\
				C006 - C022 & 8.6 & 0.093 & $-$8.96 & 1.04 & $-$0.36  & 0.05   &  1.34	& 0.668  & 510 & 539 & BS & 1.24 & 5.0  &  0.239 \\ 
				C012 - C022 & 2.9 & 0.189 & $-$8.95 & 3.05 & $-$0.33  & 0.136  &  1.90 & 2.388  & 370 & 402 & BH & 1.89 & 3.48  &  0.467 \\ 
				C012 - C018 & 2.1 & 0.21  & $-$8.18 & 3.81 & $-$0.26  & 0.135  & 1.36    & 1.488  & 660 & 684 & BS &  1.17   &  4.34  & 0.250 
			\end{tabular}
			\caption{ {\em Properties of unequal binary BS models and of the final remnant.}
          Each case is characterized by:
              $\bullet$~the compactness $C$ of the individual BSs in the binary, 
              $\bullet$~the mass-ratio $q$, 
              $\bullet$~the symmetric mass ratio $\nu$, 
              $\bullet$~the two initial centers $y_c^{(i)}$, 
              $\bullet$~the initial velocities of the boost $v_x^{(i)}$, 
              $\bullet$~the ADM mass $M_0$ of the system, and $\bullet$~the  orbital ADM angular momentum $J_0$ of the system,  after the constraint-violating transient respectively. The merger and remnant are characterized by: 
              $\bullet$~the coordinate time of contact of the two stars $t_c$ (defined as the time at which the individual Noether charge densities make contact for the first time), 
              $\bullet$~the merger retarded time (defined as the time when the maximum of the modulus of the $\Psi^{2,2}_{4}$ is produced minus the travel time to the sphere where it is measured),  $\bullet$~the type of final remnant, 
               $\bullet$~the remnant mass $M_{r}m_{b}\lambda$, 
               $\bullet$~the remnant radius $R_{N}m_{b}\lambda$ (i.e., containing $99\%$ of the Noether charge),
               and $\bullet$~the main GW frequency $M_r \omega_r^0$ in the post-merger. 
				When the final remnant is a BH, the radius is computed from the expression for Kerr BHs, $R_H = M_r(1 +  \sqrt{1 - a^2})$, where $a=J_r/M_r^2\approx0.5$  is the dimensionless spin.
			}
			\label{table2}
		\end{ruledtabular}
	\end{table*}

As mentioned, our binary initial data is only approximate, but constraint violations quickly propagate
off the grid by our evolution scheme. Hence, it makes sense to evaluate the global
characteristics of the initial data not at the initial time but instead just after
the constraint-violating transient. We therefore extract numerically the ADM mass, $M_0$, of the spacetime after the transient,
and, assuming that the mass ratio remains constant through the transient, we decompose this
mass into the constituent ``effective'' masses as
\begin{eqnarray}
\Tilde{M}_1 = \left(\frac{q}{q+1}\right)M_{0} \,,\quad \Tilde{M}_2 =\left(\frac{1}{q+1}\right)M_{0} \,.\label{massefect}
\end{eqnarray}
Notice that this calculation tacitly assumes that, even after the  constraint violation  transient (approximately) ends, stars are sufficiently separated so that GR nonlinearities are sub-leading. During this transient regime, we note that the masses of the constituent stars increase which results in a decrease in the number of orbits.

Furthermore, we can construct fitting formulae for the compactness, $C(M)$, and particle number, $N(M)$, as functions of BS mass  from the equilibrium configurations of isolated BSs. We obtain
\begin{eqnarray} 
 C(M) &\approx&  0.0157 + 0.376M - 0.3 M^2 + 0.136  M^3\nonumber \\
 &&- 0.0195 M^4\,, \label{eq:SBS_fitting_C} \\
N(M)  &\approx& -0.0187+0.6221  M   +0.3872 M^2 \label{eq:SBS_fitting_Q}\,.
\end{eqnarray}
With the above functions, one can calculate the ``effective'' Noether charges and  compactnesses of the stars in our binaries as a function of their ``effective'' masses, respectively. In Table~\ref{tableEffID}\,,  we provide this data for all configurations consider in this work and Paper~I. We also provide the relative differences between the properties of the isolated initial data and the ``effective'' ones. Comparing the total Noether charge in the system, $N_0$, with the sum of the individually calculated charges, $N(\Tilde{M}_{1})+N(\Tilde{M}_{2})$, provides a test of the consistency of this approach. As explained below in Sec.~\ref{sec:binary_toy}, the ``effective'' initial data presented here  agrees roughly with our initial data after the constraint-violating transient.
\begin{table*}
	\begin{ruledtabular}
		\begin{tabular}{c|||cc|cc|cc||cc|cc|cc}
			Binaries & $\tilde{M}_1 m_b \lambda$ & $\Delta M_1/\tilde{M}_1$ & $\tilde{C}_1$ & $\Delta C_1/\tilde{C}_1$ & $\tilde{N}_1(m_b \lambda)^2$ & $\Delta N_1/\tilde{N}_1$  & $\tilde{M}_2 m_b \lambda$ & $\Delta M_2/\tilde{M}_2$ & $\tilde{C}_2$ & $\Delta C_2/\tilde{C}_2$ & $\tilde{N}_2(m_b \lambda)^2$ & $\Delta N_2/\tilde{N}_2$   \\
			\hline\hline
			C006 - C006 & 0.13 & 0.066 & 0.061 & 0.0099 & 0.071 & 0.14 &  & & & & &\\
			C012 - C012 & 0.43 & 0.15 & 0.13 & 0.093 & 0.32 & 0.21 &  & & & & &  \\
			C018 - C018 & 1.0 & 0.22 & 0.21 & 0.14 & 1.0 & 0.29 &  & & & & & \\
			C022 - C022 & 1.6 & 0.32 & 0.28 & 0.20 & 1.9 & 0.42 &  & & & & &  \\
			C003 - C022 &  0.048 & 0.034 & 0.033 & 0.093 & 0.012 & 0.38 & 1.1 & 0.035 & 0.22 & 0.0028 & 1.2 & 0.034 \\
			C006 - C022 & 0.14 & 0.11 &  0.063 & 0.044 & 0.076 & 0.20 & 1.2 & 0.11 & 0.23 & 0.044 & 1.3 & 0.13 \\
			C012 - C022  & 0.49 & 0.25 & 0.14 & 0.16 & 0.38 & 0.32 & 1.4 & 0.24 & 0.25 & 0.14 & 1.6 & 0.32 \\
			C012 - C018  & 0.44 & 0.17 & 0.13 & 0.10 & 0.33 & 0.23 & 0.92 & 0.15 & 0.20 & 0.10 & 0.88 & 0.19
		\end{tabular}
		\caption{ {\em Effective properties of the individual stars within the binary after the
         constraint violating transient.} Tildes represent ``effective`` quantities of
         the stars in the binary.
			For the equal mass binaries of Paper~I and the unequal mass binaries studied here:
                  $\bullet$~the mass from Eq.~\eqref{massefect}, 
                  $\bullet$~the compactness $C$ from Eq.~\eqref{eq:SBS_fitting_C}, 
                  $\bullet$~the Noether charge from Eq.~\eqref{eq:SBS_fitting_Q},
                  $\bullet$~for each of these, their fractional differences, $\Delta X/X$,
                            with respect to the initial data for the isolated star used in the
                            construction of the binary.
		}
		\label{tableEffID}
	\end{ruledtabular}
\end{table*}	

\subsection{Numerical setup and analysis}
	
The computational code, generated by the \textit{Simflowny} platform~\cite{ARBONA20132321,ARBONA2018170,PALENZUELA2021107675,simflownywebpage}, runs under the SAMRAI infrastructure~\cite{Hornung:2002,Gunney:2016,samraiwebpage}, which provides parallelization and the adaptive mesh refinement~(AMR) required to resolve the different scales in the problem. We use fourth-order spatial,  finite difference operators to discretize the EKG equations, which are evolved in time using a fourth-order Runge-Kutta integrator~\cite{Palenzuela:2018sly}.
	
Our computational domain ranges within $[-264,264]^3$ and contains 8 levels of refinement. Each level has twice the resolution of its coarser parent level, achieving a resolution of $\Delta x_{8}= 0.03125$ on the finest grid. We use a Courant factor $\lambda_{c}\equiv\Delta t_{l} / \Delta x_{l} = 0.4$ on each refinement level $l$ to ensure the stability of the numerical scheme.	
	
We analyze some relevant global physical quantities from our simulations, such as the Arnowitt-Deser-Misner~(ADM) and the Komar mass, the ADM angular momentum, and the Noether charge, computed as in Ref.~\cite{bezpalen}. We focus our attention mainly on the gravitational radiation represented by the strain $h$, which is the quantity directly observable by GW detectors. We consider first the Newman-Penrose scalar $\Psi_{4}$, which can be expanded in terms of spin-weighted $s=-2$ spherical harmonics~\cite{rezbish,brugman} as
\begin{equation}
r \Psi_4 (t,r,\theta,\phi) = \sum_{l,m} \Psi_4^{l,m} (t,r) \, {}^{-2}Y_{l,m} (\theta,\phi),
\label{eq:psi4}
\end{equation}
where the coefficients $\Psi_4^{l,m}$ are extracted and calculated on spherical surfaces at different extraction radii. The relation between this scalar and the two polarizations of the strain is given by $\Psi_{4} = \ddot{h}_{+} - i\,\ddot{h}_{\times}$. The components of the strain in the time domain can be calculated by performing the inverse Fourier transform of the strain in the frequency domain, $h^{l,m}(t) \equiv {\cal F}^{-1} [{\tilde h}^{l,m}(f)],$ where a high-pass filter has been applied in the frequency domain in order to attenuate the signal with frequencies lower than the initial orbital frequency~\cite{2011CQGra..28s5015R,bezpalen}.
The instantaneous angular frequency of each GW mode can be calculated easily from $\Psi_4$ as
\begin{eqnarray}
\omega^{l,m}_{\rm{GW}} = -\frac{1}{m} \mathfrak{Im} \left(\frac{\dot{\Psi}_4^{l,m}}{\Psi_4^{l,m}}\right)~,~~~
f^{l,m}_{\rm{GW}} = \frac{\omega^{l,m}_{\rm{GW}}}{2\pi}.
\end{eqnarray}
We will refer to $\omega_{\rm{GW}}$ as the one given by the dominant mode $l=m=2$.

The mass, the angular momentum, and $\Psi_4$ are calculated on spherical surfaces at different extraction radii between $R_{\rm{ext}}=100$ and $R_{\rm{ext}}=200$, which are located far away from the sources in the wave zone.

\section{Dynamics for unequal-mass BS binaries}\label{dynamics}
We have evolved four unequal mass binary BS cases, $\{$C003-C022, C006-C022, C012-C022, C012-C018$\}$, covering mass ratios $q\equiv m_{1}/m_{2}$ roughly between $2$ and $23$. Additionally, we have studied a variation of the most extreme case, C003-C022A, in which the heavier BS has been transformed into an anti-BS. In what follows, we describe first qualitatively the dynamics for all the cases  and then analyze the GWs produced by these mergers in the next section.

\subsection{Binary dynamics in the inspiral}
\begin{figure*}	
	\centering
	\includegraphics[width=1.0\textwidth]{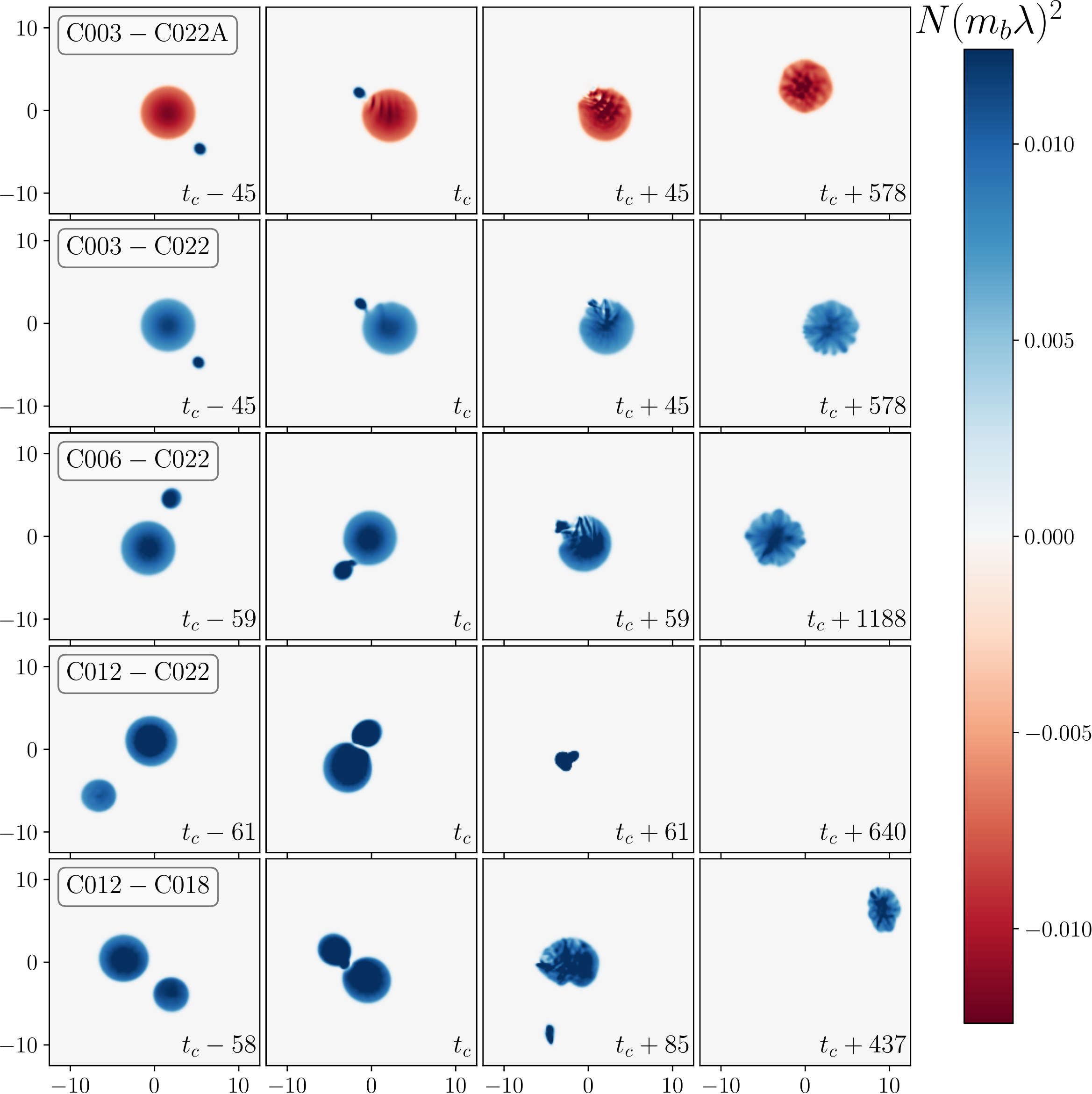}
	\caption{ {\em Dynamics of the Noether charge}. Noether charge densities in the equatorial plane $(z=0)$ at representative times. Each row represents one of the cases (from top to bottom): $\{$C003-C022A, C003-C022, C006-C022, C012-C022, C012-C018$\}$. The first column illustrates a time roughly one orbit before the contact time $t_{c}$ (defined as the time at which the individual Noether charge densities make contact for the first time), the second column occurs at contact time, the third is roughly an orbit after the contact time (except for the C012-C018 case, in order to visualize the ejected  blob), and the fourth one occurs at the end of our simulations. Note that the final remnant for C012-C022 is a rotating BH which quickly engulfs the surrounding scalar field (i.e., not visible on this natural scale).}
	\label{fig:noether_snapshots}
\end{figure*}

\begin{figure*}
	\centering
	\includegraphics[width=1\textwidth]{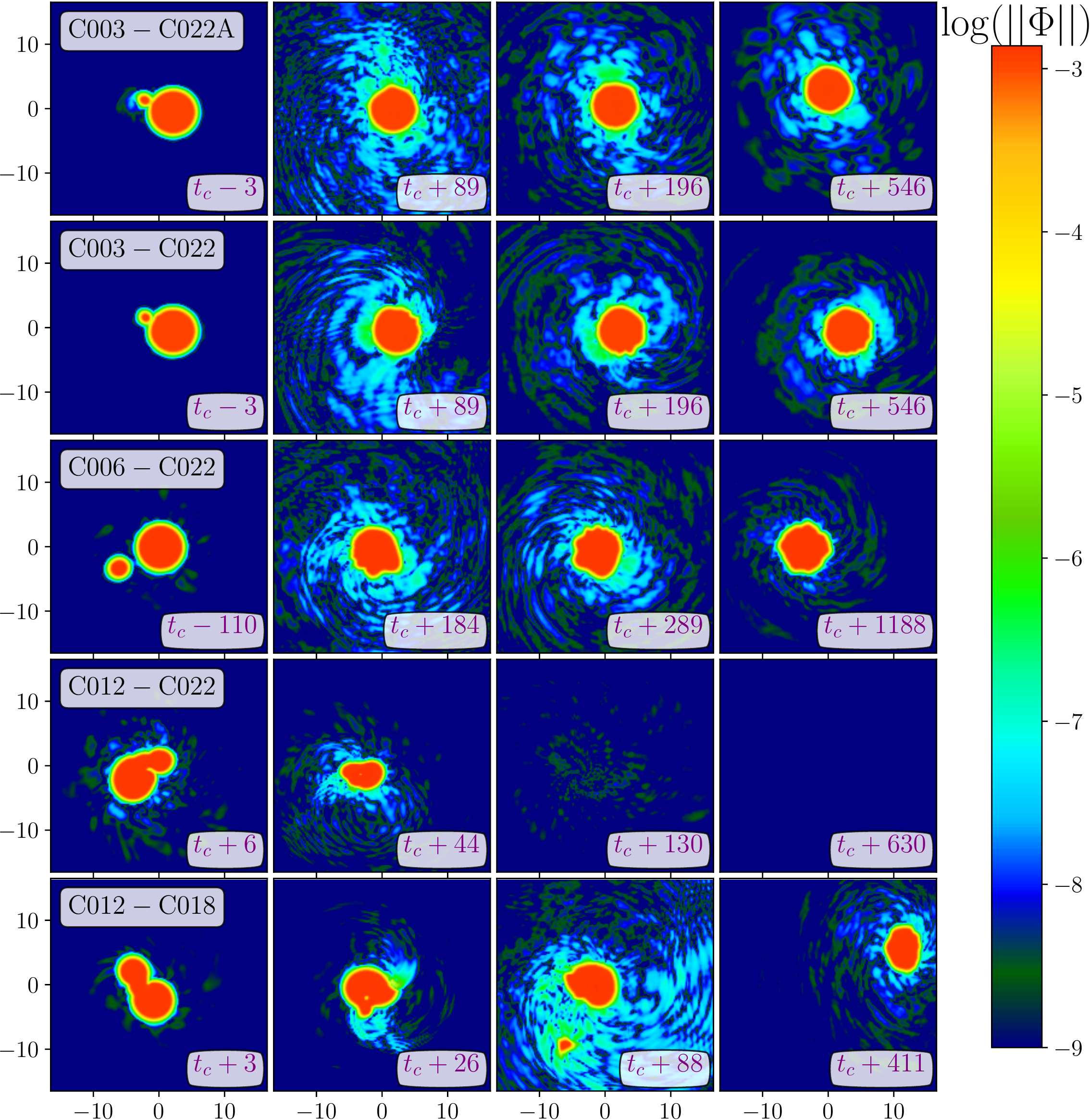}
	\caption{ {\em Dynamics of the scalar field}. Logarithm of the  modulus of the scalar field $||\Phi||$ in the equatorial plane $(z=0)$, at representative times. Each row represents the unequal cases considered. Notice that there is only some scalar emission around the contact time $t_{c}$ (defined as the time at which the individual Noether charge densities make contact for the first time), suggesting that the final object is an almost stationary BS (except for the simulation on the third row, in which the remnant is a spinning BH).}
	\label{fig:phi2_snapshots}
\end{figure*}

We display some representative snapshots along the equatorial plane
to characterize the dynamics of these binary evolutions. In particular,
the Noether charge densities in Fig.~\ref{fig:noether_snapshots}
show the dynamics of the condensed bosons, whereas the scalar field norm in Fig.~\ref{fig:phi2_snapshots} shows the dynamics of the scalar
field generally.

The binaries in C003-C022A and C003-C022 complete five full orbits before colliding, C006-C022 and C012-C018 complete three orbits, and C012-C022 performs just two. While such a short inspiral limits their use for guiding templates, the inspiral is long enough for constraint violations resulting from the construction of the initial data to propagate away.

During the inspiral, the spacetime curvature is dominated mainly
by the heavier BS, which moves in a spiral trajectory very close to the origin (i.e., see the leftmost column of Fig.~\ref{fig:noether_snapshots}),  while the lighter one induces a  perturbation orbiting around the most massive object.
This effect is especially pronounced in the four most unequal mass cases
in which the heavier BS accounts for at least $75\%$ of the binary mass.
During the inspiral, the scalar field constituting each star has no significant
overlap (see the first column of Fig.~\ref{fig:phi2_snapshots}), and therefore
nonlinear scalar interactions only play a significant role inside the stars.
Roughly speaking, the BSs behave then like point particles with moderate deviations produced by the tidal deformations.
As the mass ratio approaches unity, the binary behaves similarly to the equal-mass cases
of Paper~1. In particular, C012-C018 with $q=2.1$ resembles those equal-mass cases.

The aforementioned deviations due to tidal deformations  can be estimated by looking at the quadrupole-moment tensor $Q_{ab}^{(i)}$ of the $i$-th object induced by the tidal-field tensor $G_{ab}^{(j)}$ produced by the $j$-th object ($i,j=1,2$)~\cite{Flanagan:2007ix,PoissonWill},
\begin{equation}
	Q_{ab}^{(i)}=\lambda_i G_{ab}^{(j)}\sim \lambda_i \frac{m_j}{r^3}\,, \label{inducedQ}
\end{equation}
where $r$ is the orbital distance and $\lambda_i=\frac{2}{3}m_i^5 k^{(i)}_{\rm tidal}$ is the tidal Love number of the $i$-th object, with $k^{(i)}_{\rm tidal}$ being its dimensionless counterpart. Hence, the dimensionless quadrupole moment, $\bar Q_i=|Q_{ab}^{(i)}|/m_i^3$, reads 
\begin{align}
	\bar Q_1 &\sim k^{(1)}_{\rm tidal} \frac{q^2}{(1+q)^3}\frac{M_0^3}{r^3} \,,\\
	\bar Q_2 &\sim k^{(2)}_{\rm tidal} \frac{q}{(1+q)^3}\frac{M_0^3}{r^3} \,,
\end{align}
where $M_0=m_1+m_2$ is the binary total mass. In the large mass-ratio limit, $q\gg1$, the tidally-induced quadrupole moments of the primary and of the secondary are suppressed by a factor $q^{-1}$ and $q^{-2}$, respectively. For example, for a fixed value of $k^{(i)}_{\rm tidal},$ the tidally-induced quadrupole moment of the primary for $q=23$ is suppressed by a factor $3$ relative to $q=1$, whereas that of the secondary is even a factor $\sim100$ smaller. Overall, tidal effects on the secondary object are less relevant than those on the primary.

\subsection{Final fate of the binary merger} \label{sec:binary_toy}

If the system is sufficiently massive such that the remaining mass after merger exceeds the maximum stable BS mass (i.e., $M_r \geq M_{\rm max} \approx 1.85/(m_b \lambda)$), one expects the system to collapse to a remnant BH. If instead the total mass is below this threshold, a remnant BS is expected.
In the latter case, the possibility of forming a rotating BS should be considered. At least two conditions appear to be required for such formation:
(i)~because rotating BSs have quantized angular momentum, binaries need to have angular momentum at the point of contact  at least  slightly larger than or equal to the first discrete level of the rotating star,\,\footnote{This argument excludes some exotic possibility in which, say, GWs with some opposite angular momentum are radiated copiously until the remnant achieves the sufficient amount of angular momentum.} and (ii)~the rotating solution to which the remnant might settle must be stable.

\begin{figure}
	\centering
	\includegraphics[width=0.5\textwidth]{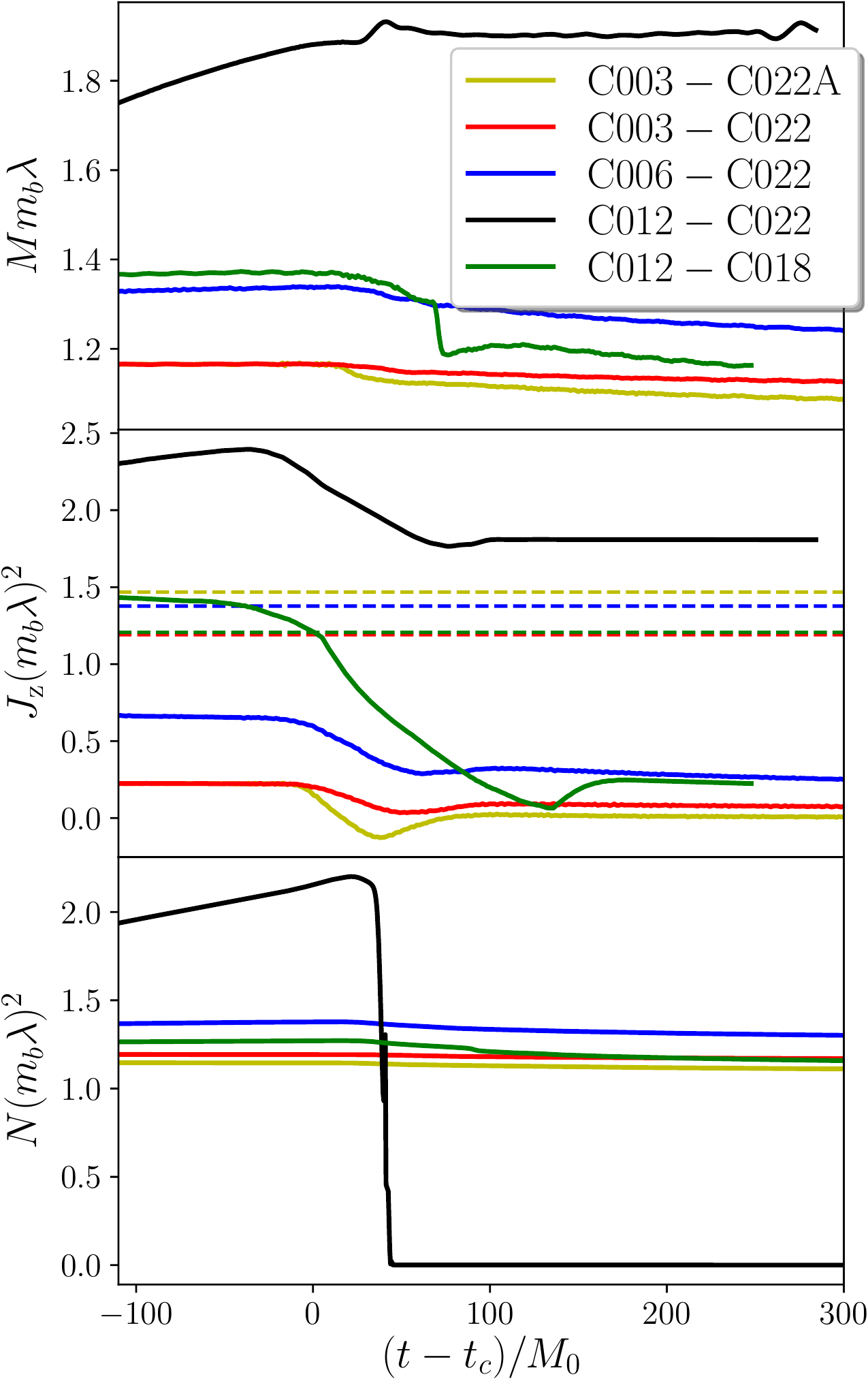}
	\caption{ {\em Global quantities}.  \rm{ADM} mass (\textbf{top}) , angular momentum $J_{\rm{z}}$ (\textbf{middle}) and Noether charge (\textbf{bottom}) as functions of time.
		The time has been shifted such that contact time happens at $t=0$ and rescaled with the initial total mass $M_0$ of each binary.         Horizontal dashed lines in the middle panel indicate the angular momentum of the ground state rotating BS corresponding to the Noether charge measured at the contact time. Notice that the angular momentum decreases monotonically (although slowly) after merger for all remnants except for that of C012-C022 which collapses to a BH.
		This decrease in angular momentum along with the fact that these binaries (except C012-C022 which collapses to a BH) have less angular momentum than any rotating BS with the same Noether charge support our claim that all non-collapsed cases settle to a nonrotating BS.
		The non-monotonic, brief drops in the mass and angular momentum plots for the C003-C022A and C012-C018
		cases correspond to the passage of transients across the extraction surface on which mass and
		angular momentum are calculated (the retarded time is used). The Noether charge is computed as
		a volume integral and therefore less subject to such errors.
	}
	\label{fig:MNJ}
\end{figure}

Once the stars contact each other, one expects scalar field interactions to produce additional attractive forces that accelerate the merger (see the discussion of the effective force with just a massive potential in Appendix~B of~\cite{pale1}). The newly formed, rotating object is initially largely nonaxisymmetric, and,
even by the end of our simulations, the remnant is a highly perturbed BS
(see the rightmost column of Fig.~\ref{fig:noether_snapshots}). Some general features of the dynamics can be found in certain global quantities (mass, Noether charge, and  angular momentum) which are displayed in Fig.~\ref{fig:MNJ}.

\begin{figure*}[th]
	\begin{tabular}{cc}
		\includegraphics[width=1\textwidth]{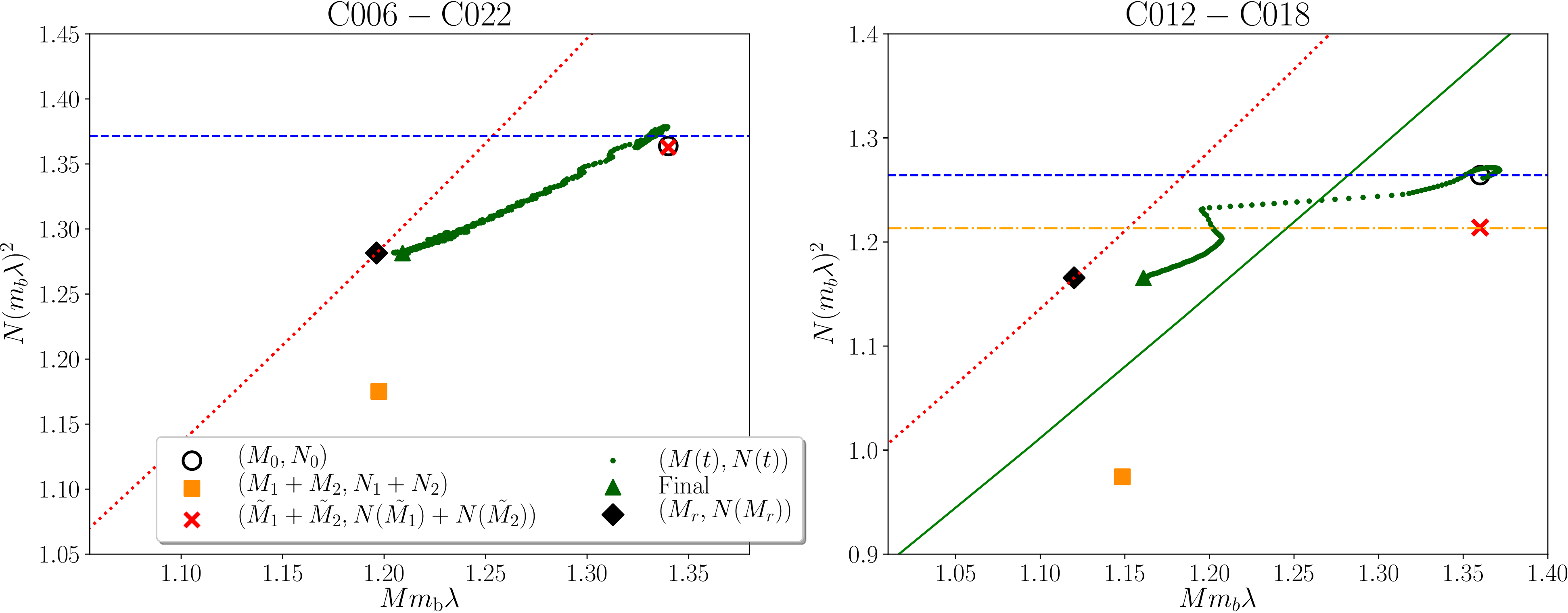}
	\end{tabular}
	\caption{{\em Mass-Noether charge phase space}.
		Evolution of the total mass and charge of the system for two representative cases: C006-C012 (\textbf{left}) and  C012-C018 (\textbf{right}) which ejects a blob of scalar field after merger.
		The orange square for each case represents the value of $(M_{1}+M_{2},N_{1}+N_{2})$ obtained from the individual stars as listed in Table~\ref{isolatedbs}.
		The black, open circles are the masses and charges, $(M_0,N_0)$, obtained from the numerical evolution just after the transient.
		The red crosses are the ``effective'' values $(\Tilde{M}_1+\Tilde{M}_2,N(\Tilde{M_1})+N(\Tilde{M_2}))$,  as explained in Sec.~\ref{subsec:IDBS}.  The fact that the red crosses and black circles are close to each other supports the validity of this approach. The green dots trace the numerical evolution by showing the extracted quantities $(M(t),N(t))$.
		The values characterizing the final state, $(M_r,N_r)$, of the simulation are represented by a green triangle. The black diamond is the point, $(M_r,N(M_r))$, with the same mass as the green triangle but with the charge obtained from the fit in Eq.~\eqref{eq:SBS_fitting_Q}. If one assumes that the remnant is a BS that relaxes only via emission of gravitational waves, one
		obtains a horizontal phase space trajectory (blue dashed line) through the initial data  (namely the black circle here).  The family of nonrotating BSs
		given by Eq.~\eqref{eq:SBS_fitting_Q} are plotted with a (red dotted curve). Only the case on the right has angular momentum comparable to the first rotating solution, and so for this case we also show the family of $k=1$ rotating BS configurations for our same potential from Ref.~\cite{2021PhRvD.103d4022S} with a solid green curve. Because $J=kN$ for such rotating BSs, we also show the value of $N$ corresponding to the angular momentum of the binary at contact time with a horizontal, yellow dot-dashed line.
	}
	\label{fig:par_space}
\end{figure*}

The mass and the Noether charge are unambiguously defined global quantities, in contrast to the radius of the star. In the case of a complex field, the $U(1)$ symmetry, which ensures the conservation of the Noether charge, significantly restricts the ways in which the remnants might relax. Fig.~\ref{fig:par_space} shows the mass-Noether charge phase space for two representative cases C006-C012 and C012-C018. Here, we present several estimates of the initial and final data along with families of isolated BSs, to facilitate the understanding of the relaxation of the remnant.

The orange squares indicate the simplest estimate of the initial data, $(M_{1}+M_{2},N_{1}+N_{2})$,
obtained by adding the properties of the isolated BSs used to construct the binary. These two squares fall far from our two other estimates of the initial data. In particular, the total mass     and Noether charge measured by the numerics after the transient is shown in black circles. We then construct the ``effective'' initial data (red crosses) by decomposing the numerically obtained     total mass via Eq.~\eqref{massefect} and computing the charge of each BSs from these individual masses~(with Eq.~\eqref{eq:SBS_fitting_Q} in Sec.~\ref{subsec:IDBS}).

We further note that, due to the nonlinearity of the function $N(M)$, some amount of scalar and/or GW emission is needed during the merger in order for the remnant to settle into either a static or rotating configuration. If the remnant is assumed to be a BS that relaxes only by the emission of GWs, namely no emission of scalar field to infinity, the evolutionary path of the binary would follow a horizontal line in the $N$-$M$ phase space (blue dashed line on Fig.~\ref{fig:par_space}), ultimately settling into the remnant BS occurring at the intersection with the family of nonrotating BSs given by Eq.~\eqref{eq:SBS_fitting_Q} (red dotted line).
Our simulations indicate emission of scalar field, in addition to GWs, a process known as ``gravitational cooling''~\cite{Seidel:1993zk,Guzman:2006yc}.
Indeed, the  path of the numerical evolution (green dots) indicates that the dynamics are driving each system toward a stationary BS (red-dotted line). Although most of these BS mergers ended before the remnant fully relaxed to stationarity, we have established for C003-C022 and C012-C012 that the point $(M_{r},N_{r})$ (where $N_r\equiv N(t_\mathrm{end})$) indeed lies on the  isolated BS curve. However, the near constancy of the Noether charge in the late postmerger (Fig.~\ref{fig:MNJ}) and the close approach of the final simulation to the isolated BS curve (Fig.~\ref{fig:par_space}) both indicate that the mergers that do not collapse are forming a stable, nonrotating, solitonic BS.

If the late stage evolution is dominated by GW emission (since most of the ambient scalar field has already been radiated), then we would expect the final object to be that represented by the black solid diamond, $(M_r,N(M_r))$, itself a stationary BS, because the Noether charge would not be changing.

\begin{figure*}[th]
	\begin{tabular}{cc}
		\includegraphics[width=1\textwidth]{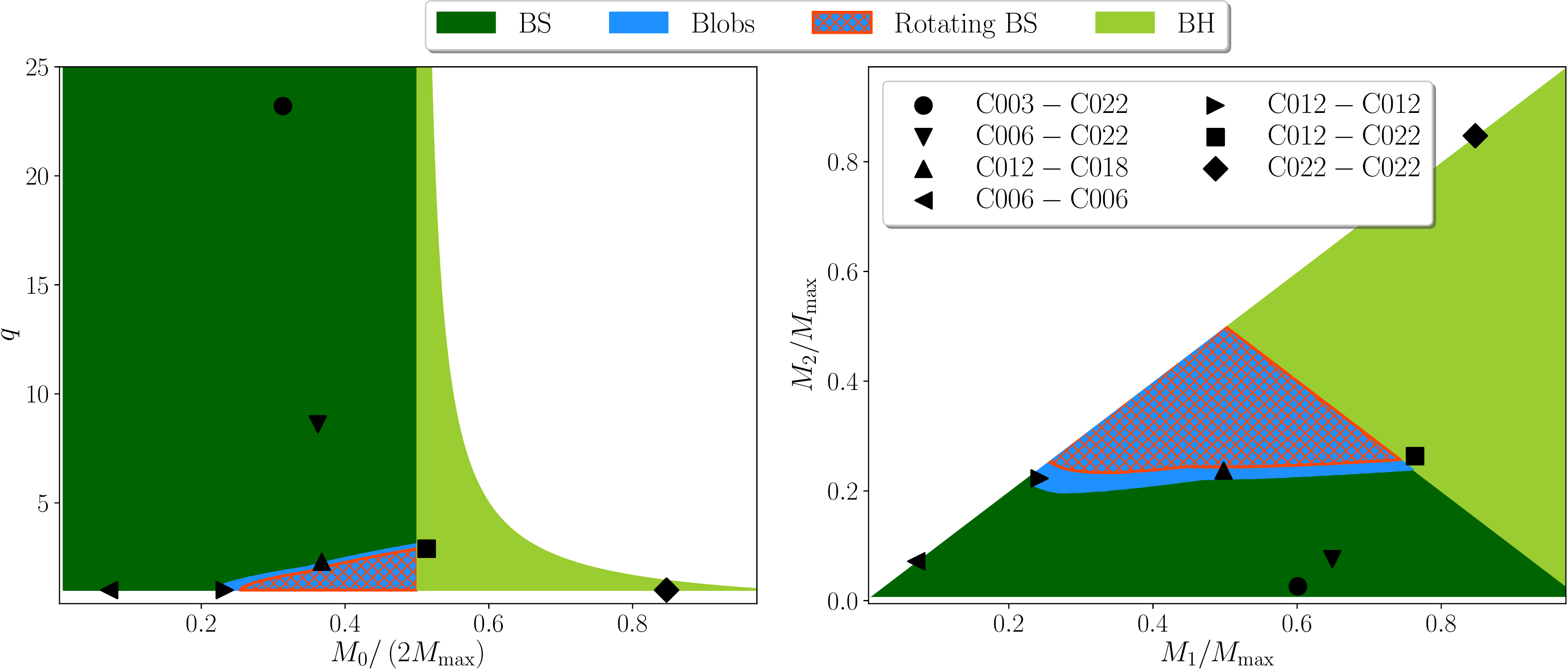}
	\end{tabular}
	\caption{{\em Scenario classification}. The parameter space of solutions represented in two different ways:
		({\bf{left}}) mass ratio versus total mass, and ({\bf{right}}) individual mass of one star versus the other. The outcomes of our simulations are consistent with $M_{\max}$ being the simple delineator for the BS/BH nature of the remnant. The blue region encloses configurations that satisfy the angular momentum requirement of Eq.~\eqref{eq:angmomcondition} (informed by our evolutions that produce blobs), and thus we expect either blob ejection or the formation of a rotating boson star. The red hashed region is a subset of this region which we have not sampled, but where 
rotating BSs may form. 
	}
	\label{fig:scenarios}
\end{figure*}

An important unresolved question is whether a merger of two BSs can produce a rotating BS. The stability of rotating, solitonic BSs has been studied recently. First, rotating BSs without scalar self-interactions were found to be unstable due to a non-axisymmetric instability~\cite{2019PhRvL.123v1101S}. However, a subsequent study showed that this instability was quenched for the solitonic model of the potential~\cite{2021PhRvD.103d4022S} (see also Ref.~\cite{Dmitriev:2021utv}) if $M>0.13/(m_b \lambda )$, for the value $\sigma_0=0.05$ considered here. Without stability, one would not expect formation of such configurations from a merger.

Rotating BSs have quantized angular momentum, $J=k N$ for some integer $k$, and one can calculate the function $N(M)$ for the $k=1$ family of rotating BSs following Ref.~\cite{2021PhRvD.103d4022S} (see also~\cite{Kleihaus:2005me}). We display this family of solutions as a green solid curve in the right panel of Fig.~\ref{fig:par_space} (case C012-C018) because this binary has angular momentum close to this first quantized level. Actually, only two cases among those studied in this work and Paper~I (i.e., C012-C018 and C012-C012) are close to satisfying the quantization condition, namely that the angular momentum is greater than or equal to the Noether charge at the time of contact. In neither of these two cases do we find a rotating remnant, and the angular momentum is primarily reduced through emission of scalar ``blobs.''

The case C012-C018 is shown in the right panel of Fig.~\ref{fig:par_space}. We display the Noether charge equal to the binary's angular momentum at the time of contact with the horizontal, yellow dot-dashed line. However, as shown in the figure, the point of intersection of the dynamical path of the binary, $(M(t),N(t))$ shown in green dots, with the curve indicating the $k=1$ family of rotating BSs (solid green curve) occurs above this yellow line. Because these rotating solutions have angular momentum equal to their charge, the evolution lacks sufficient angular momentum to form the rotating BS indicated by this point of intersection.

In an effort to understand the configuration space of binaries in terms of possible endstates,
in particular including formation of a rotating remnant or a blob,
we parameterize the quantization condition. We first compute  a Keplerian estimate of  the angular momentum  either at the time of first contact $R_{c}=C_1/M_1+C_2/M_2$ or when the binary reaches  the innermost stable circular orbit, $R_{\rm ISCO}=6M_0$, whichever occurs first.
We then correct this estimate by including the relativistic effects of strong gravity. Due to the pre-contact scalar emission, the total Noether charge in the binary at the  point of contact will be slightly smaller than the initial one. In addition, we have observed blob emission in both cases where the total charge of the binary is slightly higher than $J_c$. We incorporate these two effects in our quantization condition for  rotating boson stars, $J \geq N$, by introducing two new parameters $\{ e_{N} \,, e_{J} \}$ in the following way
\begin{equation}
	\frac{J_{c,K} (1+e_{J})}{N(M_1)+N(M_2)}> 1 + e_{N},
	\label{eq:angmomcondition}
\end{equation}
where $J_{c,K}$ is the Keplerian estimate of the angular momentum at the contact time, $e_{N}$ estimates either the amount of Noether charge radiated during the merger $(e_N>0)$ or the difference between the critical angular momentum and the charge at the point of contact that allows for blob emission $(e_N<0)$. Finally, $e_{J}$ accounts for general relativistic corrections to the Keplerian angular momentum calculation.

We use the above cases to estimate the value for the parameters $e_{J}, e_{N}$. To estimate $e_J$, we compute the differences between the Keplerian estimate of the angular momentum at contact time and the numerical value, obtaining $\sim 25\%$ in scenarios where we observe blob formation: C012-C012 and C012-C018. In the low-mass regime, where solitonic BSs are in the weak-field regime, we expect that $e_{J} \to 0$. Thus, we linearly interpolate $e_J$ between 0 and 0.25 for $M_0/2 \in [M_{\rm min},M_{\rm C012}]$ and take the constant value $e_J=0.25$ up to $M_0=M_{\rm max}$. Due to the initial data constraint violation, we cannot estimate reliably how much of the Noether charge is emitted before contact. For the sake of argument, we take $e_N=0.01$ in this case, delineating a subset of the parameter space where the strict form of the quantization condition is satisfied and where rotating remnants may form. In addition, we require that the remnant has surpassed the threshold mass estimate from Ref.~\cite{2021PhRvD.103d4022S}.

In the two cases where the blobs are observed, one finds $e_N=\{-0.05,-0.02\}$ for C012-C018 and C012-C012, respectively. Thus, taking $e_N=-0.05$ would encompass both scenarios where blobs are found and indicate the part of the parameter space where one can expect blobs generically and even possibly rotating remnants (more restrictive condition). We sketch the configuration space for these mergers  in Fig.~\ref{fig:scenarios} in two ways: the  left panel plots the mass ratio versus the total mass, $(q, M)$, whereas the right panel shows the space spanned by individual masses $(M_{1}m_{b}\lambda, M_{2}m_{b}\lambda)$. Solutions exist only for binaries constructed with stable BSs, $M_i<M_\mathrm{max}$, with regions outside this indicated in white. For binaries with $M_1+M_2<M_\mathrm{max},$ the formation of a rotating BS appears possible for the binaries that do not collapse to a BH and possess angular momentum satisfying Eq.~\eqref{eq:angmomcondition}, although we have not observed such formation (red hashed region).\footnote{Because the maximal mass of rotating BSs is larger than that for non-rotating BSs, a priori, even binaries with total mass slightly higher than the maximum mass for static stars, $M_\mathrm{max}$, could allow for the formation of a rotating remnant. Note, however, that the effective mass of C012-C022 is slightly	larger than the static maximum mass $M_\mathrm{max}$  and the configuration collapses to BH. Whether this also happens for $q\to1$ is an open question.} The set  where we expect blob emission based on the results of C012-C018 and C012-C012 cases (blue region) has a red hashed region as its subset.  Note that lacking an understanding of the physics of the blob formation, the blue region should serve only for illustrative purposes.   Both of these regions are determined approximately and require more simulations in order to understand their precise extent.

One expects qualitatively similar behavior near $M_0 \approx M_\mathrm{max}$ in the small $\lambda$ 
regime ($\lambda \ll 1$). In contrast, when $M_0 
\to M_{\rm min}$  BSs behave as thick-walled
Q-balls (where ``Q-balls'' \cite{Coleman:1985ki} refers to the flatspace limit of solitonic BSs)~\cite{Boskovic:2021nfs}, we can study the quantization condition~\eqref{eq:angmomcondition} in detail in this regime. We consider an equal-mass ($q=1$) binary with $N \approx \lambda M$ in which  the objects collide at $R_c$ (for $q=1$ this happens when $C<0.17$).
Taking $e_J \approx e_N \approx 0$ (in $m_b=\lambda^{-1}$ units) and setting the angular velocity to the Keplerian estimate, it can
be shown with some algebra that Eq.~\eqref{eq:angmomcondition} becomes
\begin{equation}
	C< \frac{M^2}{4\lambda^2}  \,.
\label{eq:qball}
\end{equation}
Thus, for sufficiently small $\lambda$ (approximately an order of magnitude smaller than the value in this work $\lambda=0.25$), the quantization condition will be satisfied. This simple expression does not change parametrically when a more precise description of the Q-balls is used \cite{Boskovic:2021nfs}. Although rotating Q-ball solutions have been constructed \cite{Kleihaus:2005me}, the non-axisymmetric instability~(NAI) probably prevents one from dynamically forming, based on the results of Ref.~\cite{2021PhRvD.103d4022S}. Whether in those cases blobs form or the non-axisymmetric instability would kick in is an open question.

To conclude,  we cannot rule out the formation of a rotating BS with the solitonic potential although none has been formed. In any case, our parameter space analysis indicates that the initial conditions would need significant tuning, which may require more accurate initial data. Even in those cases where the formation of rotating BS might be feasible, as suggested in Paper~I, the organization of the bosonic field into a rotating star from the very nonlinear merger may be too difficult, particularly because the rotating BS necessarily has a toroidal energy density~\footnote{Rotating Proca stars instead have a spheroidal energy density and yet none of these have been formed from a merger either~\cite{PhysRevD.99.024017}.}\,.

\subsection{Remnants: Scalar clouds, blobs, and kicks}

Another interesting possibility is the formation of a stable scalar cloud surrounding the remnant rotating BH. A necessary condition for the scalar field to remain around a spinning
BH is the saturation of the superradiant condition~\cite{Brito:2015oca}. In particular, this condition is saturated when the phase oscillation frequency of the scalar field, $\omega$, 
is  synchronized with the angular frequency of the BH, $\Omega_H$,  such that $\Omega_H = \omega/m$ for some integer $m$.
Synchronized scalar clouds were not found originally from the mergers of Proca stars~\cite{PhysRevD.99.024017}, but more recent and detailed  equal-mass binary simulations of non-solitonic bosonic stars showed long-lived, scalar hair around a rotating horizon for a small range of the initial angular momentum~\cite{Sanchis-Gual:2020mzb}. 

\begin{figure}
	\centering
	\includegraphics[width=0.45\textwidth]{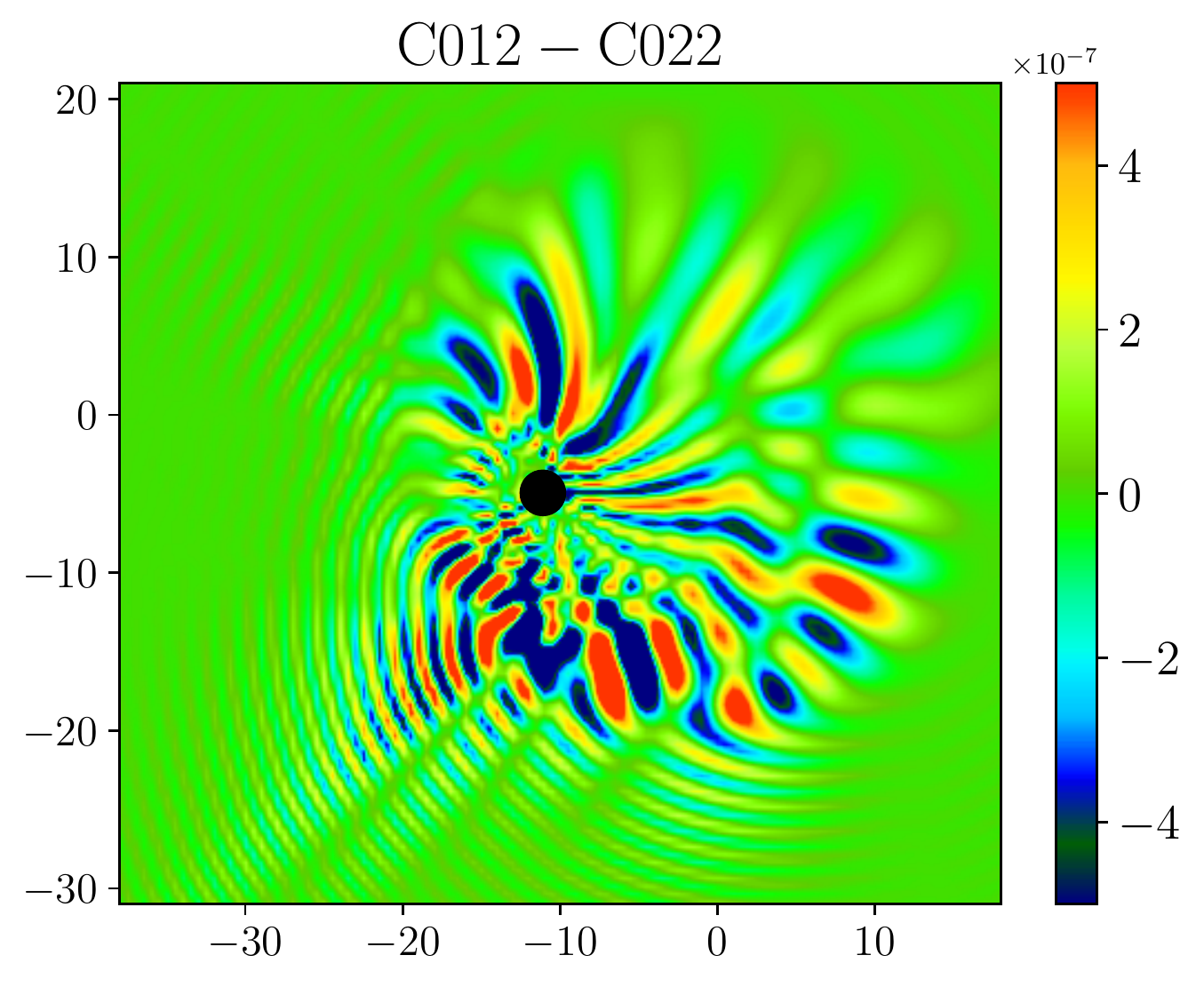} \\
	\includegraphics[width=0.45\textwidth]{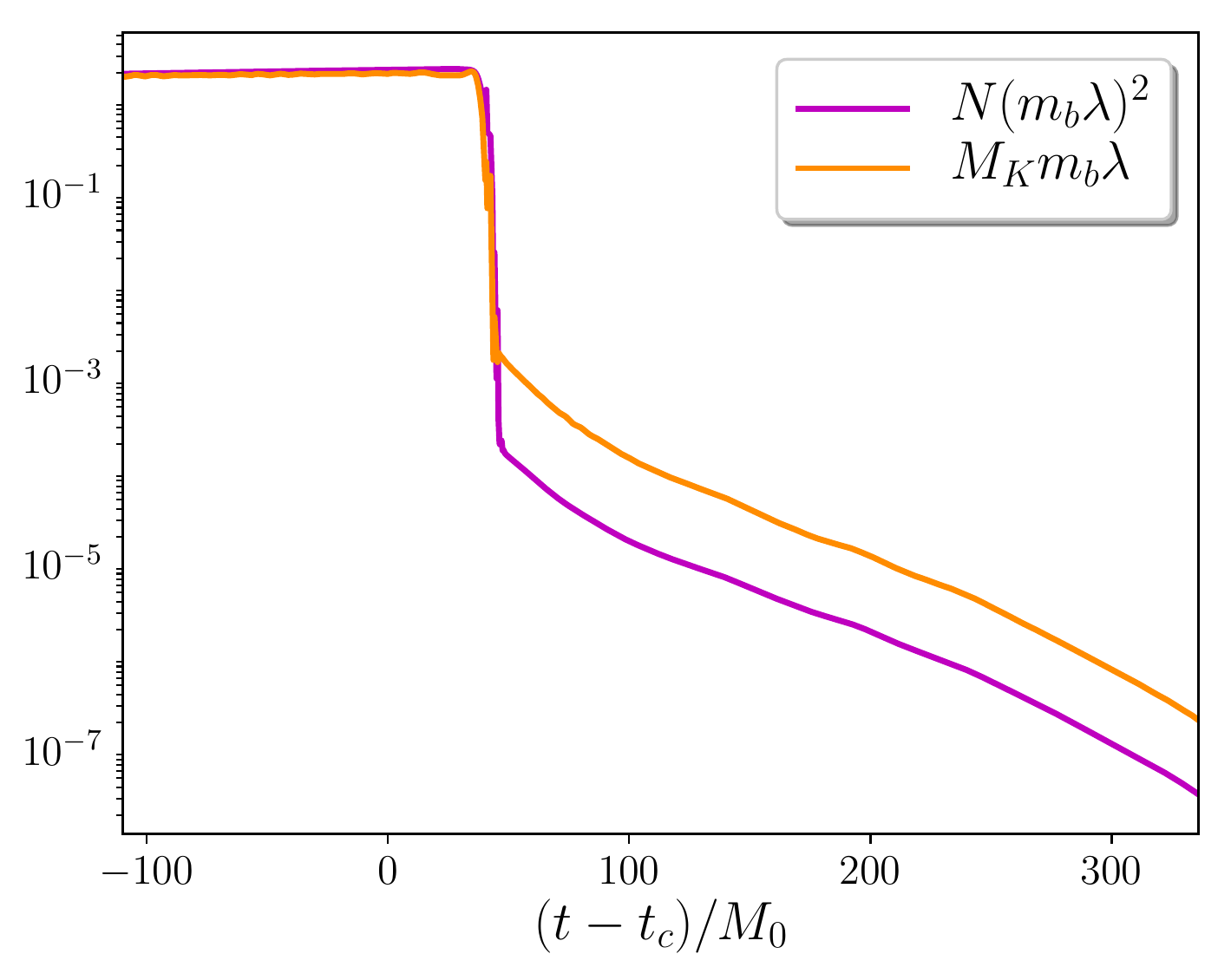}
	\caption{ {\em Details of the late-time behavior of case C012-C022}. (\textbf{Top}) Snapshot  of the real part of the scalar field at the latest time. Some scalar field remains around the BH (whose apparent horizon is represented by the black sphere located at the center of the plot) with a large $m$ configuration.
    The amplitude of the scalar field decays quickly during the timescale of our simulations and does not reach the stationary state expected for a single-mode synchronized scalar cloud.
	(\textbf{Bottom}) Noether charge and Komar mass (formulated only in terms of the stress-energy tensor and hence measures only the scalar energy and not that of the BH) as a function of time for the same case. After the sudden drop at the merger, both quantities decay exponentially as the scalar field falls into the BH.
   }
	\label{fig:logphi_NMK}
\end{figure}

We examine the case C012-C022 to determine whether any scalar field remains after the remnant has collapsed to a BH. Visualizing the scalar field and its associated Noether charged density reveals
no significant remaining scalar field (see the fourth column of the  the C012-C022 case of Figs.~\ref{fig:noether_snapshots} and~\ref{fig:phi2_snapshots}). Furthermore, one sees the total
Noether charge drop quickly to zero after merger in Fig.~\ref{fig:MNJ} and the bottom panel of Fig.~\ref{fig:logphi_NMK}.

To evaluate the possibility of formation of a synchronized scalar cloud, we calculate the oscillation frequency of the scalar field for a numerical comparison of the synchronization condition $\Omega_H=\omega/m$. We note first that the final BH has mass $M=1.89/(m_b\lambda)$ and angular momentum $J_{z}=1.92/(m_b\lambda)^2$, leading to a dimensionless spin $a=J_{z}/M^2=0.537$. The radius and angular frequency of the BH can be computed from expressions for Kerr-Schild BHs as $R_H = M (1 + \sqrt{1 - a^2}) \approx 3.5/(m_b\lambda)$ and $\Omega_H = a/ (2 R_H) \approx 0.077 m_b\lambda$. 
We Fourier transform the scalar field at an arbitrary point outside the BH but where the scalar field is well above any numerical noise (roughly a distance of $12 /(m_b\lambda)$ from
the BH, as in the top panel of Fig.~\ref{fig:logphi_NMK}). We find a frequency  $\omega \approx 0.6 m_b\lambda$ which, with the synchronization condition and the estimate of the BH rotation $\Omega_H$, implies an azimuthal quantum number $m\geq 8$. Interestingly, as shown in the top panel of Fig.~\ref{fig:logphi_NMK}, the real, and similarly the imaginary~(not shown), components of the late-time scalar field configuration outside the BH resemble the high $m$-number structure of a stationary cloud. However, the amplitude of the scalar field is decreasing fast, consistent with the decrease in both the Komar mass and Noether charge, displayed in the bottom panel of the same figure.

Previous studies suggest that initial data might need to be fine tuned in order to form a stationary configuration, unless such a configuration is a dynamical attractor as in the case of the superradiant instability~\cite{Brito:2015oca}. 
It is worth noticing that Ref.~\cite{Sanchis-Gual:2020mzb} found Proca clouds with $m$ as high as $6$, but in the vector case the superradiant instability develops much faster than in the scalar case at hand.
In the small $m_b M$ limit, the instability time scale of scalar fields is longer than that of vectors by a factor $(m_b M)^{-2}\gg1$~\cite{Brito:2015oca}. 
Furthermore, the instability is very suppressed for large azimuthal number $m$. The imaginary part of the fundamental $(l,m)$ mode for a perturbation with spin $s$ ($s=0,1$ for scalars and vectors, respectively) depends on the factor $\beta_{ls}=\left[\frac{(l-s)!(l+s)!}{(2l)!(2l+1)!!}\right]^2$~\cite{Starobinskij2,Brito:2015oca}. This dependence is responsible for a suppression of the superradiant instability time scale when $m\gg1$. Indeed, for the most relevant $l=m\gg1$, and focusing only on the $m$-dependence, the instability time scale reads
\begin{eqnarray}
 \tau = \omega_I^{-1} \sim m^{2m} (m_b M)^{-4m-5+2s}\,,
\end{eqnarray}
which quickly becomes extremely large as $m$ increases. Note that the fastest growing superradiantly unstable mode in the scalar case ($l=m=1$) has an instability time scale at least $\tau\sim 10^5 M$ for a nearly extremal BH. This is already much longer than the time scale of our simulations, and it becomes much longer for $m\gg1$  modes and moderately spinning BHs. This discussion strongly suggests that it is unlikely that an $m=8$ superradiant cloud could form dynamically over the limited time scale of our simulations.

It might be possible that mergers leading to smaller oscillation frequencies of the remnant scalar field or with higher initial angular momentum (so that the final BH is rapidly rotating) are more likely to produce clouds. Either of these conditions would lead to a smaller required $m$, but may limit the parameter space of cloud-generating solutions. Clearly, more work is needed to answer this question.

We now consider the ejection of scalar blobs. As previously explained, the case C012-C018 is the only one with contact angular momentum close to that of the first quantized spinning BS configuration (namely, $J_z \gtrsim N$) that does not collapse to a BH. 
Instead, whether the spheroidal energy density formed in the merger somehow prevents the configuration from relaxing to the toroidal shape of the rotating BS or not, the system relaxes instead to a nonrotating BS. To do so, the system must shed its angular momentum.

In this case, the excess angular momentum is emitted in the form of a blob of scalar field that is ejected from the remnant soon after the merger (see the bottom row of Figs.~\ref{fig:noether_snapshots} and~\ref{fig:phi2_snapshots}). This blob travels outward on the grid, and its passage across the spherical surface (i.e., around $(t-t_c)/M_0\approx 100$)  at which the system mass and angular momentum are computed disrupts the assumptions of the calculation, seen as non-monotonicity in the global quantities shown in Fig.~\ref{fig:MNJ}.

Using the values before and after the drop in mass, we can estimate the blob's mass as $M_\mathrm{blob}\approx 0.12$. Despite the blob containing only a small fraction of the total mass, it carries a significant fraction of the total angular momentum  due to its large velocity, $v_\mathrm{blob} \approx 0.5$ directed nearly tangentially away from the remnant  and its distance from the center of mass when ejected, $L\approx 7$. Indeed, using the same simple estimate for the angular momentum as in Paper~I, we obtain $J_{z} \approx M_\mathrm{blob} v_\mathrm{blob} L \approx 0.4$, which is roughly equal to the sharp decrease of angular momentum observed in the middle panel of Fig.~\ref{fig:MNJ}. On the time scale of our simulation, the blob appears bounded. In fact, the blob satisfies the stability condition (in $m_b=\lambda^{-1}$ units) $\lambda  M_\mathrm{blob} <  N_\mathrm{blob},$ with $N_\mathrm{blob} \approx 0.05/(m_b \lambda)^2 $ (see Ref.~\cite{Boskovic:2021nfs} and references therein for  a discussion of the stability regimes of solitonic BSs).

In addition to the unequal mass case C012-C018 presented here, the ejection of condensed scalar field was observed in two equal-mass BS binary simulations, one in Ref.~\cite{bezpalen} and the other in Paper~I. In those two cases, the symmetry of the binary resulted in two, identical blobs propagating along opposite directions. That three different studies found blob ejection
suggests that such ejection might be typical in solitonic BS binaries under certain conditions. 

The ejection of the blobs has important implications for the astrophysics of BS mergers should such systems (or similar systems such as axion stars) actually exist in nature. In contrast to
the equal-mass case that ejects two blobs in opposite directions, the ejection of a single blob generates a kick on the remnant. For binaries with large enough mass ratios, the kick can be large; large even compared to the superkicks of binary BHs (which are as large as a ${\rm few}\times 10^3\,{\rm km/s}$~\cite{Boyle:2007sz,Campanelli:2007cga,Gonzalez:2007hi,Lousto:2011kp}) and larger than the typical escape velocities of galaxies and of globular clusters (which are of ${\cal O}(10^2-10^3)\,{\rm km/s}$ and ${\cal O}(10)\,{\rm km/s}$, respectively).
For example, the linear momentum of the blob shown in the C012-C018 simulation is roughly $M_{\rm blob} v_{\rm blob}\approx 0.06/(m_b\lambda)$ which, by linear-momentum conservation, implies that the remnant with mass $M_r=1.17/(m_b\lambda)$ recoils with a velocity $v_r\approx 0.05\sim 1.4\times 10^4\,{\rm km/s}$ .
In practice, since the ejected scalar blobs have a sizeable mass and relativistic speed, they induce remnant kicks much more efficiently than GW emission in asymmetric binary BH systems~\cite{Boyle:2007sz,Campanelli:2007cga,Gonzalez:2007hi,Lousto:2011kp}.
These large kicks would have important implications for the merger rate of BS binaries in the universe, as they largely exceed the escape velocity from bound structures (e.g. nuclear star clusters~\cite{Antonini:2016gqe} and galaxies~\cite{Bogdanovic:2007hp}). As a result, the rate of successive generations of mergers (which is particularly important for supermassive objects, see e.g. Ref.~\cite{Barausse:2020gbp}) may be suppressed relative to the BH case. Moreover, ```stray'' BSs moving at high speeds may be present in the intergalactic medium as a result of ejections from the host galaxies.

Finally, one might be tempted to associate this disruption and blob ejection to the nonaxisymmetric instability present in some rotating BSs~\cite{2019PhRvL.123v1101S}. However, a recent study shows that the NAI should be quenched for the solitonic potential for sufficiently compact BSs~\cite{2021PhRvD.103d4022S}, which suggests that the NAI is not the cause of blob ejection.

\subsection{Collision of boson and anti-boson stars}

The use of BSs as a model of compact objects allows easily for the study of anti-stars~\cite{Dupourque:2021poh}. Here, the scalar field represents its own anti-particle simply by changing the sign of its phase oscillation, $\omega \rightarrow -\omega$ (this transformation also switches the sign of  the Noether charge associated with the BS). The case C003-C022A represents a BS with a small ``antimatter star'' in a merging binary. The evolutions of C003-C022A and C003-C022 differ only once the stars make contact, at which point the antistar annihilates. In particular, the interaction of the oppositely oscillating complex field annihilates such that the remaining field lacks the harmonic oscillation in time and the Noether charge density vanishes. Therefore, this scalar field interaction breaks the coherence of the BS solution, the dispersive nature of the scalar field dominates over the attraction of gravity, and the unbound scalar field is radiated to infinity.

The small difference between the final and  initial mass of case C003-C022A, $\Delta M \approx 0.09$, accounts roughly for twice the value of the lightest BS, $M=0.0463$. This suggests that other energies, such as that of GW radiation and binding energy, change little or are otherwise very small.  This simple calculation is also consistent with the lightest star being completely annihilated during the merger. It is interesting, and a sign of the stability of the solutions under a large perturbation, that the remnant still settles into a stable BS, even after the annihilation of a significant fraction of its Noether charge. 

\section{Gravitational Wave Signal}\label{gwsect}
We now turn our attention to the analysis of the gravitational radiation produced by unequal-mass BS binaries.
\subsection{Late inspiral and merger} \label{sec:gw_later_inspiral}

\begin{figure*}
		\centering
		\includegraphics[width=1\textwidth]{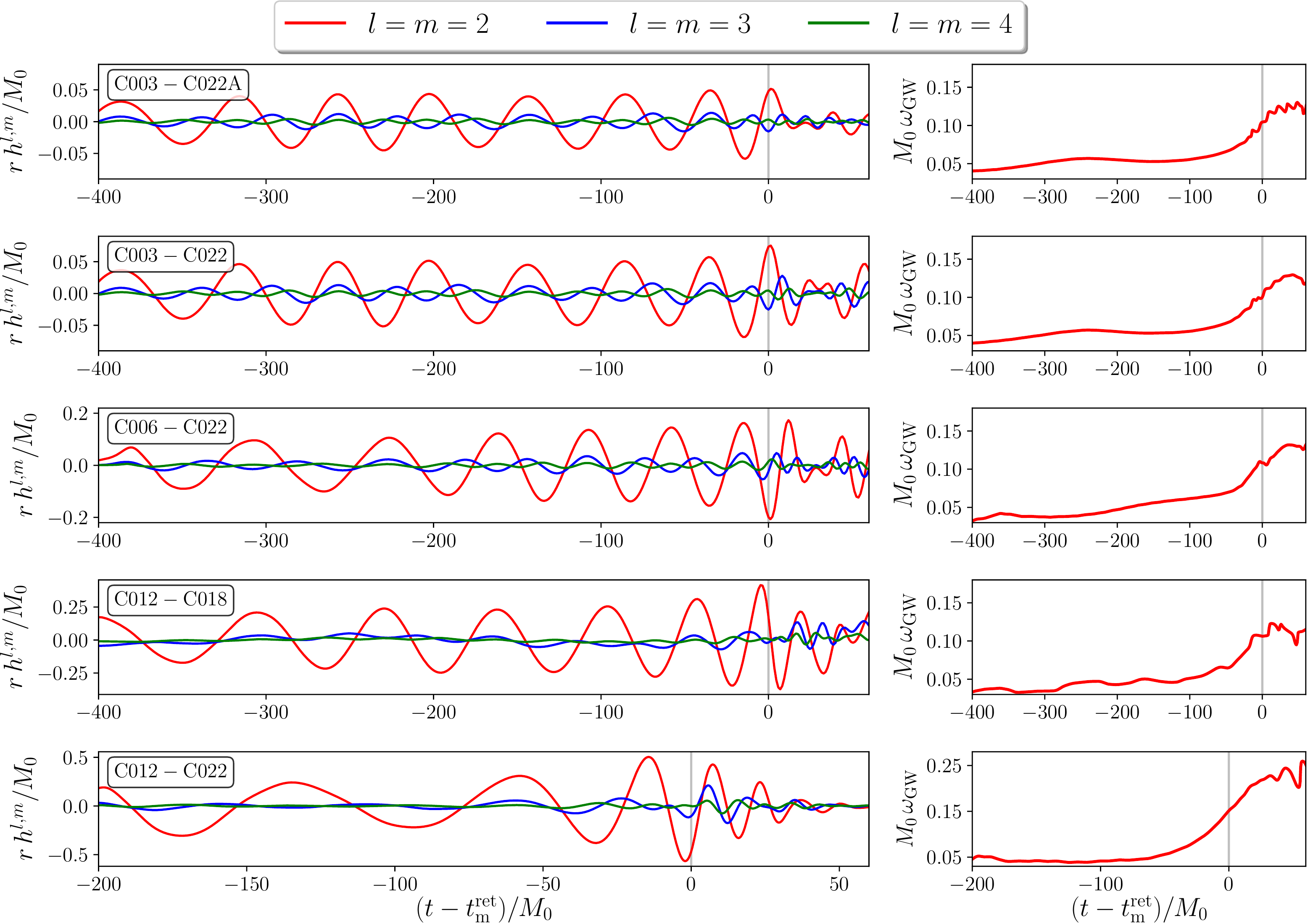}
		\caption{ {\em GWs in the coalescence}. Different modes $l=m$ of the strain as a function of time,  together with the frequency of the dominant mode $l=m=2$. Clearly, the $l=m=2$ mode is always much larger than the others, even for the largest mass ratio. The vertical, gray lines show the merger time.
        }
		\label{fig:GWall}	
\end{figure*}

Some of the most relevant $(l,m)$ modes of the gravitational radiation represented by the strain, together with the angular frequency of the $(2,2)$  mode, are displayed in Fig.~\ref{fig:GWall}. A simple inspection of these profiles already confirms that the dominant mode during the inspiral is always the $l=m=2$ for our wide range of mass ratios. As expected, mass ratios closer to unity (i.e., such as the C012-C018 case), when the mass quadrupole moment is stronger, displays a larger predominance of the $l=m=2$ mode.  On the other hand, for large mass ratios (i.e., such as the C003-C022 case), the importance of the higher-order modes increases. It is interesting to note that after the merger the amplitudes of the various modes are of the same order, without one clearly dominating over the others. 

Furthermore, as the mass ratio increases, the effects of tidal deformations on the waveform become less relevant. This can be understood as follows. A generic quadrupole-moment tensor, $Q_{ab}^{(i)}$, of the $i$-th object affects the GW phase starting at second post-Newtonian order. The extra $1/r^3$ dependence of the tidally-induced quadrupole moment [see Eq.~\eqref{inducedQ}] implies that tidal effects enter the GW phase starting at the fifth post-Newtonian order, with a phase correction~\cite{Flanagan:2007ix,PoissonWill}
\begin{equation}
 \delta \phi_{\rm tidal} =-\frac{117}{8}\frac{(1+q)^2}{q}\frac{\Lambda}{M_0^5} v^5~,
\end{equation}
where $v=(\pi M_0 f)^{1/3}$ is the orbital velocity, $f$ is the GW frequency, and $\Lambda=\frac{1}{26}((1+12/q)\lambda_1+(1+12q)\lambda_2)$ is the weighted tidal deformability.
When $q=1$, $\Lambda=(\lambda_1+\lambda_2)/2$ is simply the average of the two tidal deformability parameters. However, in the large mass-ratio limit~\cite{Pani:2019cyc}, we can write the correction as

\begin{equation}
 \delta \phi_{\rm tidal} = -k_1 \left( \frac{3}{8}  v^5 q  +\ldots \right) -k_2 \left( \frac{9}{2} v^5 \frac{1}{q^3}+\ldots\right)\,,
\end{equation}
where we include for each of the tidal terms $k_1$ and $k_2$ only the leading-order term in the $q\gg1$ expansion. The above equation shows that the tidal deformability of the primary is much more important than that of the secondary, which is suppressed by a relative factor $\sim q^{-4}$. Thus, the net contribution of the tidal deformability in the GW phase, compared to the point-particle phase, depends on two competing effects: on the one hand, less compact BSs have a large tidal Love number (see Table~\ref{isolatedbs}) but, on the other hand, for binaries with very disparate mass stars the tidal Love number of the secondary is negligible.
The quantity $\Lambda/M_0^5$, which provides a measure of the relevance of the tidal contribution compared to the leading-order point particle phase, is presented in Table~\ref{tableLOVE}. For the binary systems under consideration, the suppressing effect of large mass ratio more than compensate for the large tidal Love number of the secondary, and hence the quantity $\Lambda/M_0^5$ is larger for the smallest mass-ratio system in the catalog.

\begin{table}
 			\begin{ruledtabular}
			\begin{tabular}{l||ccccc}
				Binaries & $q$ & $k_1$ & $k_2$ & $\Lambda/M_0^5$ \\
				\hline\hline 
				C003 - C022A& 23.2 & 20 & 136494   & 0.75  \\
				C003 - C022 & 23.2 & 20 & 136494   & 0.75  \\
				C006 - C022 & 8.6  & 20 & 8420     & 0.99 \\
				C012 - C022 & 2.9  & 20 & 332      & 0.94 \\
				C012 - C018 & 2.1  & 41 & 332      & 1.79 \\
			\end{tabular}
			\caption{ {\em Tidal properties of unequal binary BS models considered in our simulations.}
                   For each binary studied here, the weighted tidal deformability $\Lambda$ is shown.
                   Despite the large tidal Love numbers in the most unequal mass binaries, the
                   binary deformability increases as $q \rightarrow 1$.
			}
			\label{tableLOVE}
		\end{ruledtabular}
\end{table}

\begin{figure}
\centering
\includegraphics[width=0.5\textwidth]{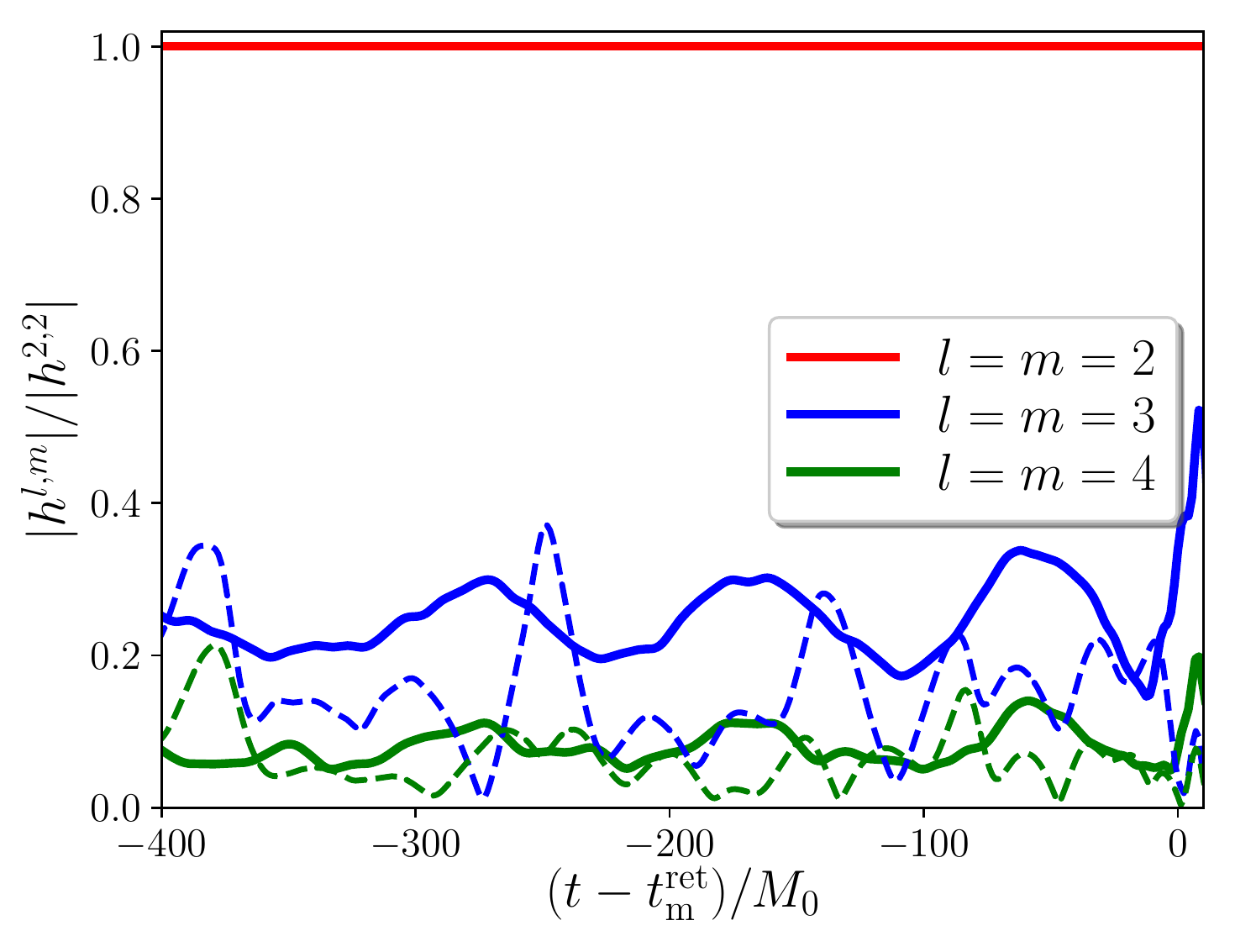}
\includegraphics[width=0.5\textwidth]{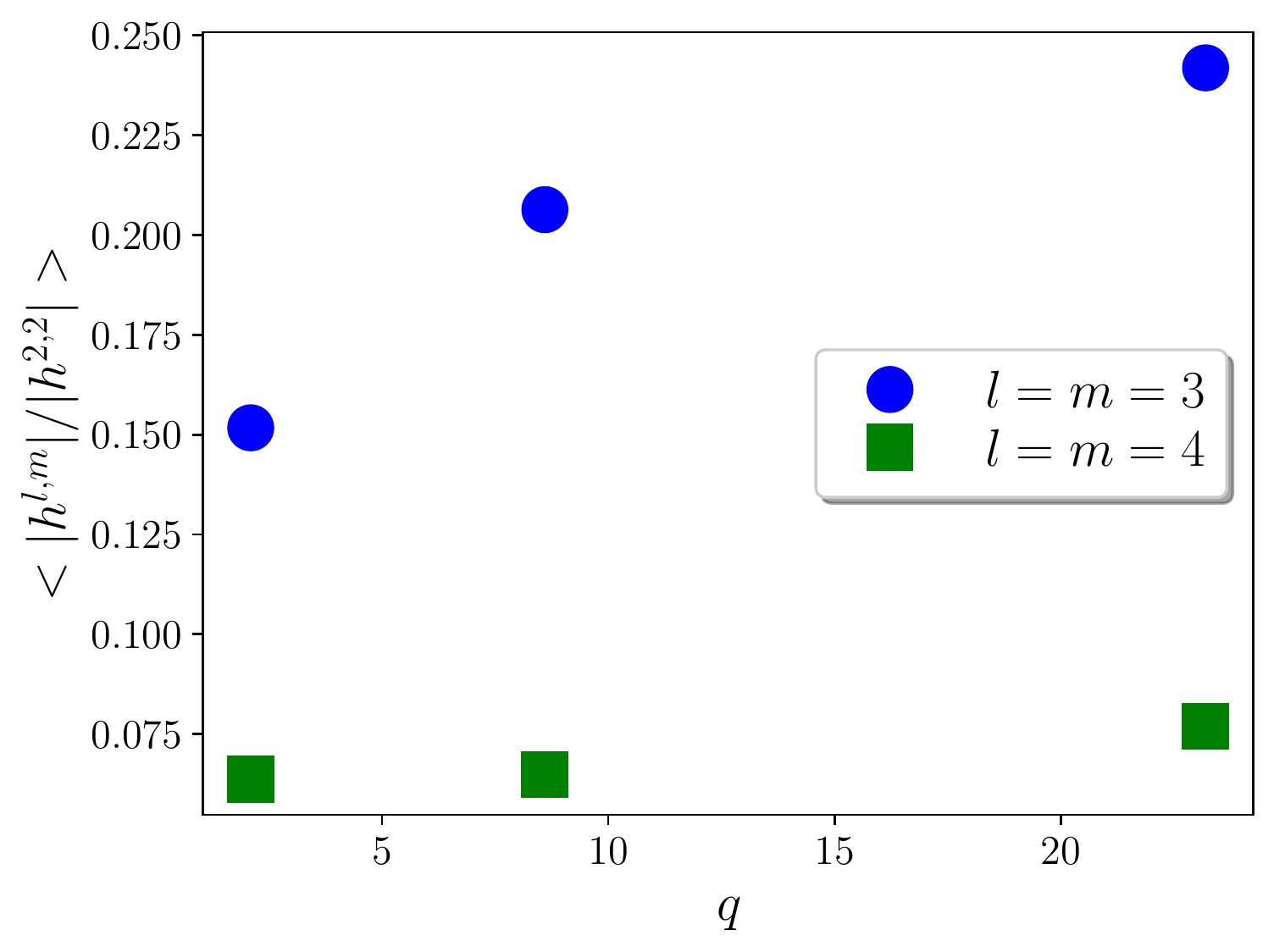}
\caption{ {\em GWs in the inspiral stage}. (\textbf{Top}) Ratio between the $l=m\ge 2$ modes of the strain and the dominant $l=m=2$ for the two most extreme cases C012-C018 ($q=2.1$, dashed lines) and C003-C022 ($q=23.2$, solid lines). (\textbf{Bottom}) The ratios averaged over the last few orbits for all the mass ratios studied. Notice that the $l=m=3$ mode increases by a factor 1.7 in this range of mass ratios, while the $l=m=4$ barely increases with mass ratio.
} 
\label{fig:hlmh22}
\end{figure}

Let us now consider the waveform's higher modes. We can characterize the effect of the mass ratio on the higher modes by examining the ratio between some relevant modes $h^{l,m}$ and the dominant $h^{2,2}$ mode. The top panel of Fig.~\ref{fig:hlmh22} displays this ratio for the two extremely unequal cases, C012-C018 with $q=2.1$ (dashed lines) and C003-C022 with $q=23.2$ (solid lines). The panel shows clearly that $h^{3,3}$ is much larger for the more unequal case while the $h^{4,4}$ case is less clear.
A more quantitative comparison can be performed by averaging the ratios $h^{l,m}/h^{2,2}$ over the last few orbits, corresponding roughly to the range of orbital frequencies $M_0 \omega_{\rm{GW}} \in(0.05,0.1)$. In this way, we are then excluding both the early inspiral, contaminated with significant constraint violations, and the post-merger phase. The bottom panel of Fig.~\ref{fig:hlmh22} displays these ratios as a function of the mass ratio. The mode $h^{3,3}$ increases relative to the dominant one by almost a factor $1.7$ when passing from $q=2$ to $q=23$, while $h^{4,4}$ barely changes. Even for our largest mass ratio, the amplitude of the $h^{3,3}$ mode is at most $25\%$ of the dominant one $h^{2,2}$. 
	
\begin{figure}
	\centering
	\includegraphics[width=0.48\textwidth]{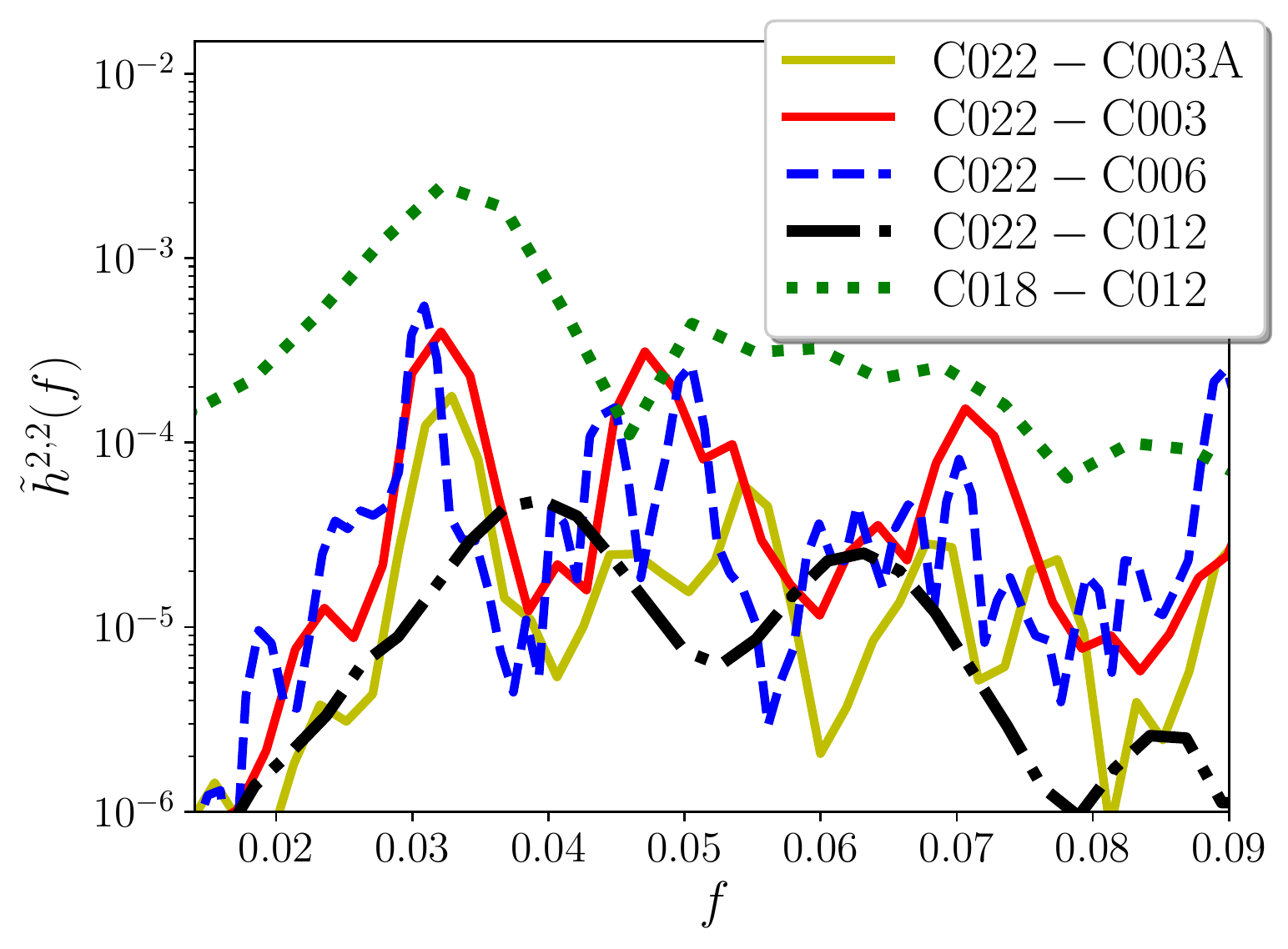} \\
	\includegraphics[width=0.48\textwidth]{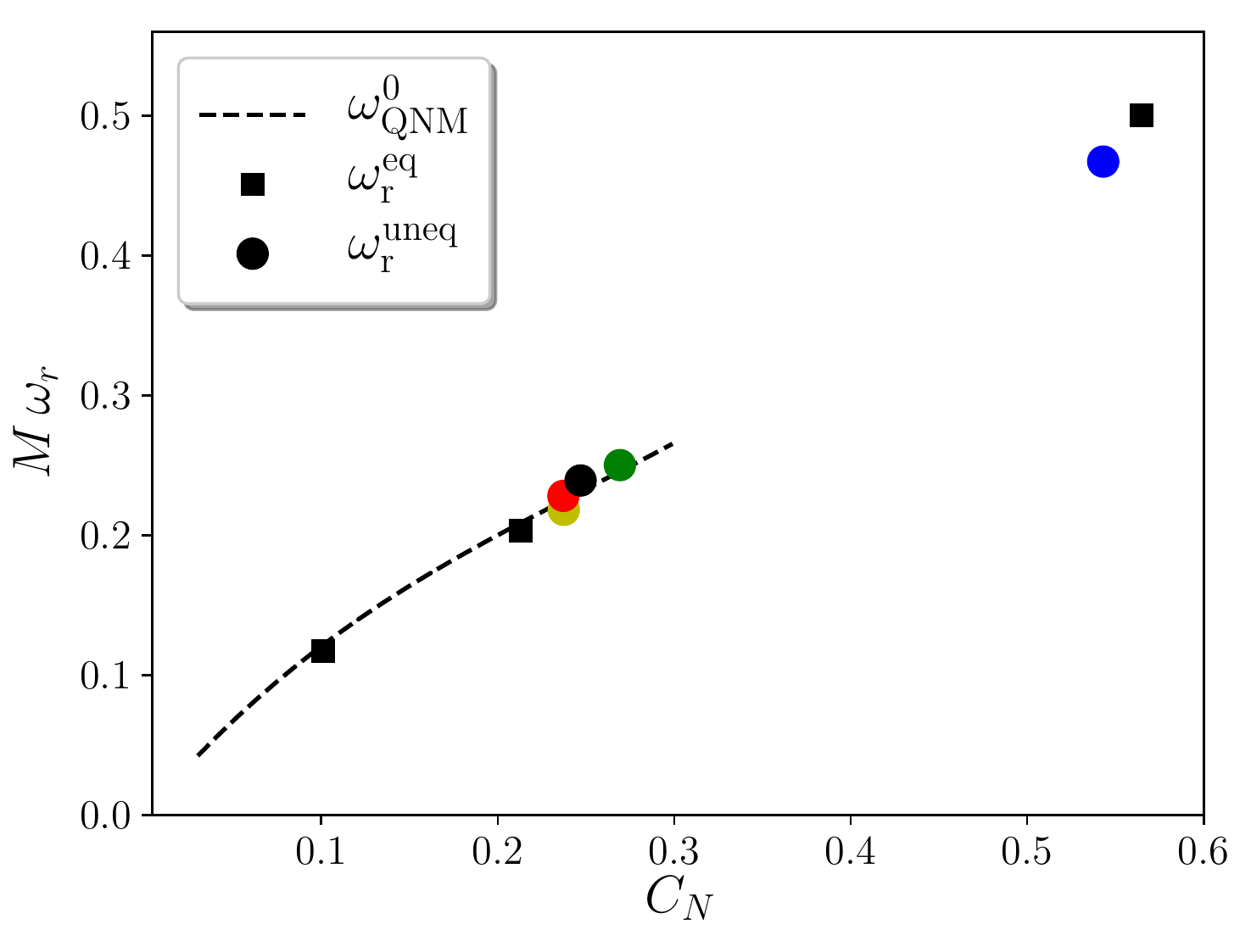}
	\caption{ {\em GWs in the post-merger stage}.  (\textbf{Top}) Power spectral density of the main mode (i.e., $l=m=2$) of the strain for the post-merger. 
(\textbf{Bottom}) Frequencies of the first peak of the $l=m=2$ mode for all the cases represented as a function of the compactness of the remnant $C_N \equiv M_{r}/R_N$ (filled circles). Reliable results from the equal-mass mergers of Paper~I are shown with filled squares. The dashed line corresponds to the lowest quasinormal mode (QNM) frequencies $\omega_\mathrm{QNM}$ of isolated BSs. The fact that the frequencies of the merger remnants are in good agreement with those of isolated BSs provides further evidence that the remnants have settled to stable BSs. Notice that we have also included the cases that collapse to rotating BHs (those with $C_N\ge0.5$).
	}
	\label{fig:freq}
\end{figure}

\subsection{Post-merger}

We analyze the post-merger frequencies of the gravitational signal of the remnant, showing the power spectral density of the dominant $l=m=2$ mode in the top panel of Fig.~\ref{fig:freq}. In the bottom panel of Fig.~\ref{fig:freq}, we display the frequency of the dominant mode for all the cases studied in this paper and Paper I, together with the fundamental mode for isolated BS stars as a function of the compactness of the remnant.

\begin{figure*}[th]
	\begin{tabular}{cc}
\includegraphics[width=1\textwidth]{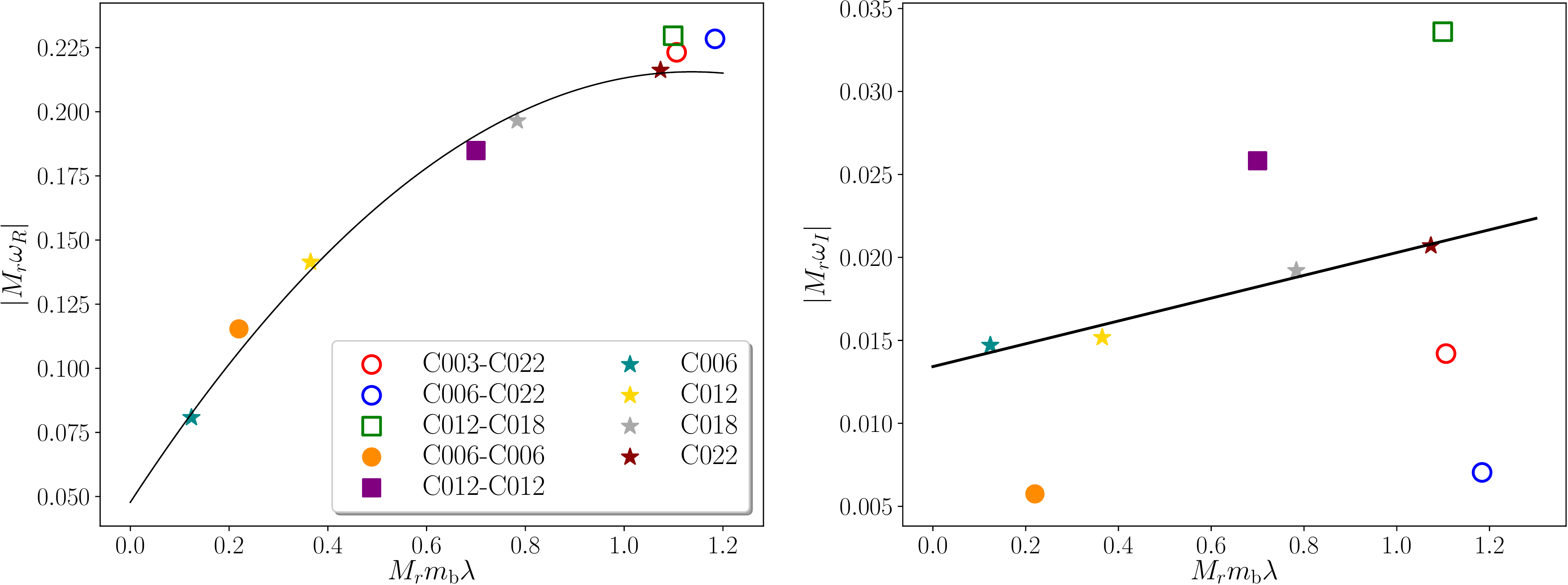}
	\end{tabular}
	\caption{{\em QNMs and post-merger spectrum.} (\textbf{Left}) Real  and (\textbf{right}) imaginary  parts of the 
             frequencies of the first peak of the $l=m=2$ mode with respect to 
             remnant mass $M_r$ for binaries that form
             a remnant BS, with the equal mass binaries
             of Paper~I (filled) and the unequal mass binaries studied here (open).
             Binaries that eject a blob are denoted with squares while circles denote 
             those that do not.
             The frequencies of the QNMs of the four isolated BSs used for initial data
             are marked with stars, and the curves (solid) represent the fits to the
             real and imaginary components of these QNM frequencies from 
             Eqs.~\eqref{eq:QNM_Iso_Re} and~\eqref{eq:QNM_Iso_Im}.
          }
	\label{fig:qnm}
\end{figure*}

An analysis from Paper~I indicates a correspondence between the frequency of the first peak with the quasi-normal mode (QNM) of isolated BSs. We scrutinize this hypothesis further by considering the post-merger behavior of all configurations from both papers as well as the QNM of isolated solitonic BSs with $C =\{0.06, 0.12, 0.18, 0.22\}$ calculated in Paper~I. We fit the spectral lines with a Lorentzian function, i.e. 
\begin{equation}
	\Tilde{h}^{2,2}(\omega) \sim  \frac{\omega_I }{ \left(\omega^2_I+(\omega-\omega_{\rm R})^2\right)} \,
\end{equation}
to determine the peak frequency of the main mode $\omega_{\rm R}$ and the inverse decay time  $\omega_I.$ In line with the discussion on the relaxation of the remnant from Sec.~\ref{sec:binary_toy}, one can   construct quadratic fits for $\omega_{\rm R}(M_r) \,, \omega_{\rm I}(M_r)$, where $M_r = M(N_r)$ [Eq.~\eqref{eq:SBS_fitting_Q}],  
\begin{eqnarray}
M_r \omega_{\rm R} &\approx & 0.05 + 0.3  M_r -0.13 (M_r)^2 \,, \label{eq:QNM_Iso_Re}  \\
M_r \omega_{\rm I} &\approx & 0.013+0.007  M_r \label{eq:QNM_Iso_Im}  \,,
\end{eqnarray}
for isolated scenarios from Paper~I. As shown in Fig.~\ref{fig:qnm} (left panel), excluding  the C018-C018 case from Paper~I where the post-merger behavior is not reliable,  real parts of the post-merger main mode frequencies agree well with the isolated solitonic BS QNM fit.

However, in the case of the imaginary frequency (see Fig.~\ref{fig:qnm}, right panel) all remnants produced in the binary coalescence have an offset with respect to the isolated QNMs. We notice that the three configurations in which blobs do not form have lower imaginary frequencies (longer decay times)  compared to the isolated configurations. In contrast, in the case of blob formation, frequencies are higher (shorter decay times) than expected from the isolated QNMs. Note the three cases with $m_b \lambda M_r \approx 1.07- 1.1 $ that have almost identical real frequencies but vastly different imaginary components. Understanding this peculiar behavior lies beyond the scope of this paper. We speculate that the  excess angular momentum requires longer decay times in contrast to the isolated configurations, except in the case of blob formation that removes  the excess (rotational) energy more efficiently than in the isolated case, thus shortening the decay time.

We have also compared the fits with the tabulated BH QNMs~\cite{rdweb}. The BH remnant from Paper~I, i.e. the C022-C022 case $(a=0.698)$, has tabulated value $(M_r \omega_R,M_r \omega_I)=$ $(0.532, -0.081)$, while we find $(0.469,-0.083)$, which is close to the time-domain fit from Paper~I where $(0.5,-0.07)$. For the C022-C012 case $(a=0.5)$, we find the tabulated value $(0.464,-0.086)$, while the fit gives $(0.475,-0.061)$. This mild discrepancy between the fit and the predicted ones for BHs may originate from the numerical precision of the ADM mass/angular momentum extraction and the fit, the presence of some remnant scalar surrounding the BH, or the fact that the frequency estimate depend on the choice of the post-merger time.
Nonetheless, the overall agreement corroborates the conclusion that the remnant is a BH.

\subsection{Solitonic BSs in the LIGO/Virgo band}

In this subsection, we quantify the difference between the GW signal expected from BS binaries and from binary BHs, focusing
on the LIGO/Virgo band. In particular, we assess whether analyzing LIGO/Virgo data with binary BH templates
can lead to missed detections or to biases on the estimate of the parameters of the source, under the assumption that the latter consists
of a BS binary.

As a preliminary test of this, we consider the BS binary waveforms extracted from the unequal-mass simulations of this paper,
focusing on the $l=m=2$ mode alone. Actually, each of these simulations can be taken to represent
a binary of any total mass, as long as frequencies and strain amplitudes
are properly rescaled, i.e. each simulation actually corresponds to a one-parameter family of systems with varying
binary mass  $M$, but with fixed dimensionless product $m_b M/\hbar$.
We choose therefore to vary $M$ in a range
likely to yield observable effects in the LIGO/Virgo frequency band, i.e. we choose
$M$ in the interval $[M_{\rm min},100] M_\odot$, where $M_{\rm min}$ is such
that the smaller progenitor is always heavier than $\sim 3 M_\odot$.
For each BS waveform obtained in this way, we rescale the (2,2)-mode strain amplitude to correspond to a fiducial luminosity distance of 400 Mpc.
(We recall that choosing a slightly different distance will simply rescale strains and signal-to-noise ratios
by a linear factor, at leading order.)
We then compare the BS signal obtained to SEOBNRv4 BH binary waveforms~\cite{Bohe:2016gbl}, as implemented in the \texttt{Pycbc} python package~\cite{alex_nitz_2021_5347736}. The component masses and luminosity distance of the BH binary waveform are chosen to match those of the BS binary, the component BH spins are set to zero,
and the initial phase and merger time are chosen so as to minimize the ``difference'' of the two signals. In particular, we minimize
the signal-to-noise ratio of the difference of the two signals,
\begin{equation}
	\rho(\Delta)=\left[4 \int \frac{|\tilde{\Delta}(f)|^2}{S_n(f)}df\right]^{1/2},
\end{equation}
with $\Delta(t)\equiv h_{\rm BS}(t)-h_{\rm BH}(t)$
the residual, i.e.  the difference between BS and BH signals (computed for optimal detector orientation
and sky position), and with a tilde denoting a Fourier transform. The (single-sided) power spectral density of the noise, $S_n$, is chosen
to be that of a single LIGO detector. More precisely, we consider both the case in
which $S_n$ corresponds to the Livingston detector in O3b~\cite{LIGOScientific:2021djp}, or
to the zero-detuning, high laser power {\it design} sensitivity curve~\cite{aligo}.
Accounting for the second LIGO interferometer and for Virgo will further increase
the signal-to-noise ratio, roughly by a factor $\lesssim\sqrt{3}$ (with the $\lesssim$ due to the fact
     that the source can only be optimally placed relative to one detector at a time,
 and that Virgo is less sensitive than LIGO in O3b).

\begin{figure}[th]
\centering
\includegraphics[width=0.5\textwidth]{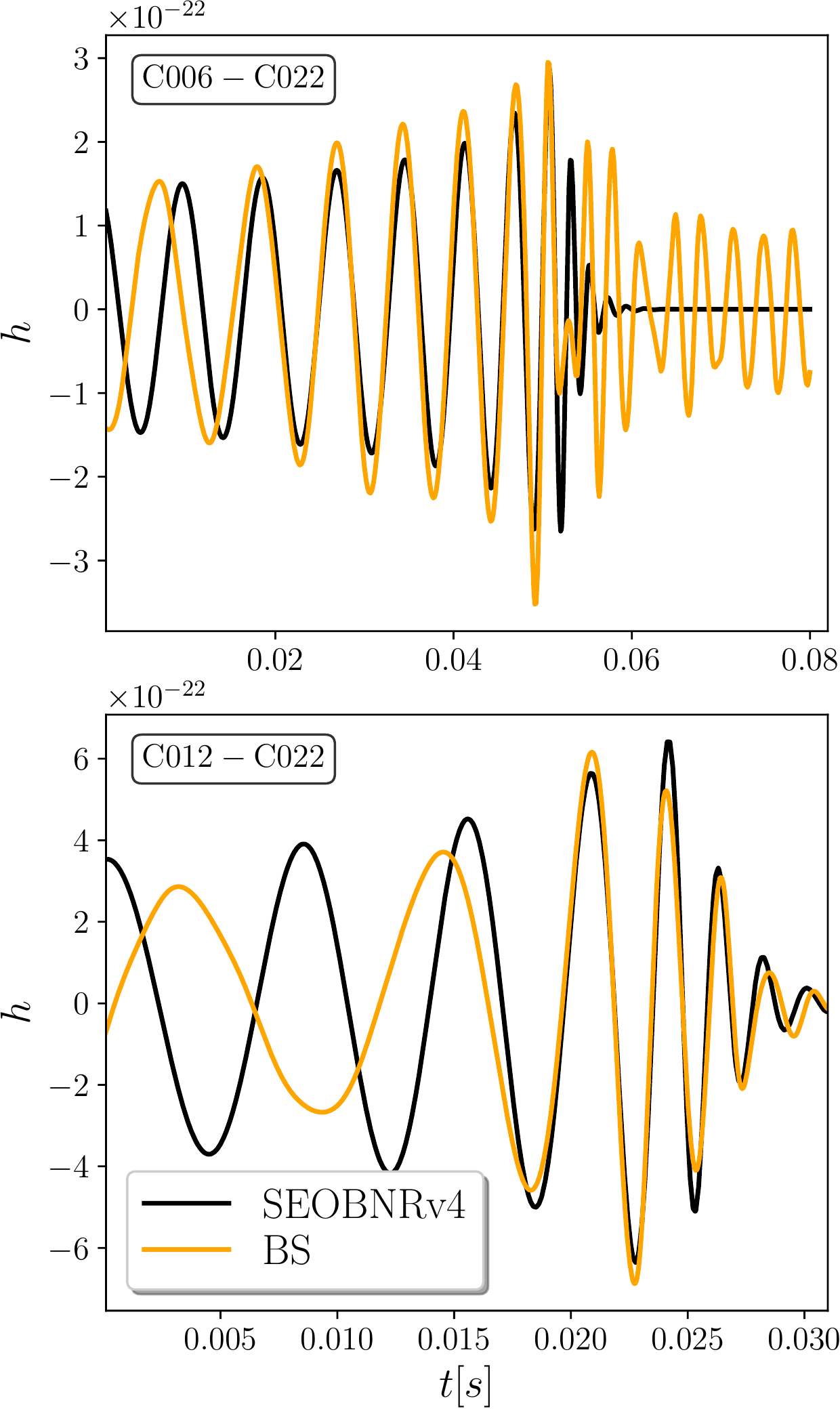}
\caption{Comparison between $(2,2)$-mode BS waveforms produced from our simulations and BH binary SEOBNRv4 templates, for a
		system leading to a BS ($q=8.6$, $M=28 M_\odot$; \textbf{top})  or a  BH ($q=2.9$, $M=30 M_\odot$; \textbf{bottom}).  The template's initial phase and merger time
		are chosen to minimize the residual signal-to-noise ratio. Both systems
		are optimally oriented at a luminosity distance of 400 Mpc, and the BH component masses and spins
		are set equal to the BS component masses and to zero, respectively.}
	\label{timeseries}
\end{figure}

Two examples of  BS binary waveforms, qualitatively 
representative of the two possible post-merger scenarios (i.e. BH or BS remnant),
are shown in Fig.~\ref{timeseries}, where they are
compared to the ``most similar'' BH binary waveforms identified with this procedure.

\begin{figure}[th]
\begin{tabular}{cc}
    	\includegraphics[width=0.5\textwidth]{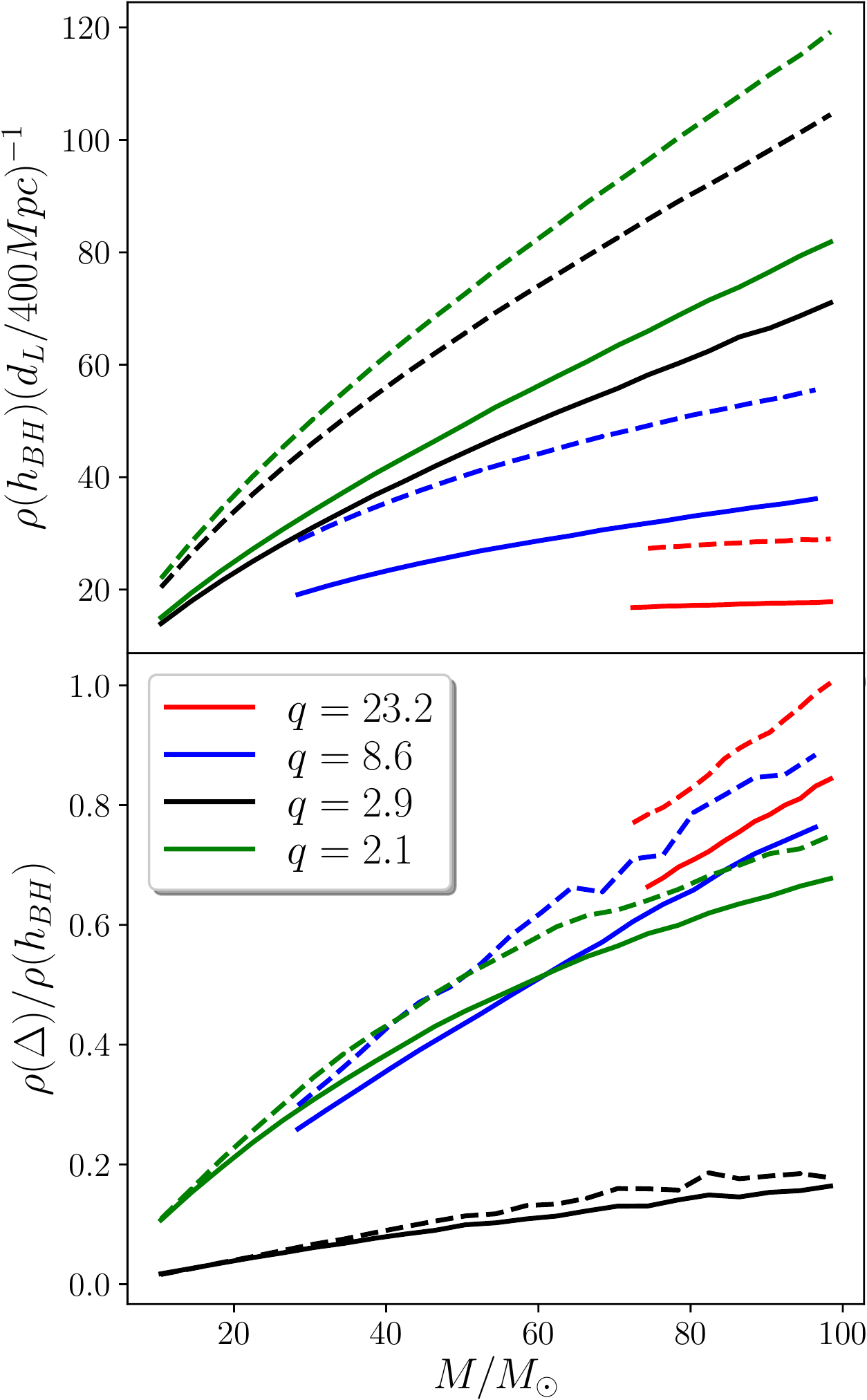}
	\end{tabular}
	\caption{Inspiral-merger-ringdown BH binary SEOBNRv4 signal-to-noise ratio $\rho(h_{\rm BH})$ (\textbf{top}) and  residual signal-to-noise ratio $\rho(\Delta)$
between BS and BH waveforms (\textbf{bottom}), as functions of the total binary mass. The residual signal-to-noise ratio is minimized over initial phase and merger time
and then normalized by $\rho(h_{\rm BH})$, for the BS waveforms extracted from our unequal mass simulations.
Both the BS and BH binaries are optimally oriented at a luminosity distance of 400 Mpc, and the BH component masses and spins
		are set respectively to the BS component masses and to zero. For both signals and templates only the $(2,2)$ mode is included.
		The signal-to-noise ratios are computed with the O3b single-detector sensitivity (solid lines), and with
		the single-detector design LIGO sensitivity in the  zero-detuning, high laser power configuration (dashed lines).
	\label{snr}}
\end{figure}

The signal-to-noise ratio $\rho(h_{\rm BH})$  of the BH binary waveform best matching each BS signal
is shown in the top panel of Fig.~\ref{snr}, as a function of $M$ and for both the O3b and design LIGO configurations. The signal-to-noise ratio is computed by using 
the aforementioned SEOBNRv4 waveforms, which (unlike our short BS signals) include inspiral, merger and ringdown. 
In the bottom panel, we show instead the residual signal-to-noise ratio $\rho(\Delta)$, minimized over initial phase and merger time
for all the simulations that we have at our disposal. 

This residual signal-to-noise ratio is computed by
comparing BH and BS waveforms that are both cut below the minimum frequency at which constraint violations are significant
in our simulations, in order to avoid biasing the comparison. 
In this way, the residual signal-to-noise ratio includes differences between BH and BS waveforms that occur in the post-merger phase and also in the late inspiral, thus including at least some contribution from tidal effects, while keeping the impact of initial constraint violations subleading~\footnote{ Improvements in the initial data would change the residual signal-to-noise ratio only marginally at high masses, while at low masses they would allow for simulating a longer portion of the inspiral phase. Our residual signal-to-noise ratios should then be regarded as lower bounds.}\,.
We then normalize  $\rho(\Delta)$ by the {\it full} inspiral-merger-ringdown
BH signal-to-noise ratio $\rho(h_{\rm BH})$, since we have no access to the full inspiral-merger-ringdown BS waveforms.
As can be seen, $\rho(\Delta)/\rho(h_{\rm{BH}})$ is
always very large and grows with $M$, as expected
because for massive binaries only the merger signal is in the band of terrestrial 
interferometers (i.e. for those binaries the BS-BH differences in the post-merger
have a larger relative impact). Also note that $\rho(\Delta)/\rho(h_{\rm{BH}})$ is 
smallest (although still quite significant) in the case where the remnant is a BH $(q=2.9)$. Again, this is expected: The collision of two non-rotating BHs with $q=2.9$ leads to a rotating remnant with $a \approx 0.52$ \cite{Hofmann:2016yih}, close to the value of the BH spin produced by the BS binary in our simulation ($a \approx 0.5$). 
As such, the differences in the merger-ringdown, where most of the in-band power resides (at least for moderately high masses),
are small, c.f. e.g. Fig.~\ref{timeseries}.

As a rough rule of thumb, residual signal-to-noise ratios
$\rho(\Delta)\lesssim 8/\sqrt{3}\sim 5$ may
allow for claiming a BS binary detection (as opposed to a BH binary one), provided
that an accurate determination of the component masses and spins is available (e.g. thanks to a long inspiral).
In the absence of a sufficiently long detected inspiral,
large residual signal-to-noise ratios may merely lead to biases in the estimation of the parameters of the source
(i.e. one could mistake a BS post-merger signal for a BH ringdown with remnant mass different from the actual one, and/or non-zero spin),
or even missed detections.
As can be seen from  Fig.~\ref{snr}, this second  possibility 
seems the most likely 
at high masses, for which most of the inspiral is out of band and 
 BH templates miss most of the signal's power for binaries producing a BS remnant.
Whether this leads to a bias on the recovered parameters or just a missed detection
should be ascertained by considering templates with varying BH progenitor spins. However,
given the long duration of the BS post-merger signal (c.f. e.g. Fig.~\ref{timeseries}), it seems unlikely that
it can be detected by any one BH template, i.e. we expect mainly missed detections at high $M$, at least
for second-generation detectors and for systems that lead to a BS remnant.
For systems that instead lead to BH formation (e.g.
the $q=2.9$ case in Fig.~\ref{snr}),
using BH templates may simply produce a bias on the parameter estimation. 

The situation will be more favorable for
third-generation interferometers~\cite{Kalogera:2021bya} such as the Einstein Telescope or Cosmic Explorer,
which will observe many more inspiral cycles.
Not only will this allow for a better measurement of progenitor masses and spins (which
will reduce degeneracies when comparing the post-merger signal to BH templates), but it
may also allow for measuring the tidal Love number in the late inspiral~\cite{Cardoso:2017cfl,Pacilio:2020jza,Pacilio:2021jmq}. This will provide additional hints
on the BS versus BH nature of the system. 
 We will explore the discovery space of these detectors, and at the same time refine
our analysis, in future work.

\section{Conclusions}\label{conclu}

The coalescence of BSs allows us to study not only the binary dynamics of one of the most viable and better motivated models of
ECOs, but also the two-body problem in General Relativity for large mass ratios. The soft dependence of the BS radius with its mass, at least for the solitonic potential used here, facilitates the numerical simulations of binaries with very different compactness, as compared to the more challenging case of asymmetric BH binaries.
Taking advantage of this feature of solitonic BSs, we have studied numerically the coalescence of unequal-mass binaries with mass ratios ranging between 2 and 23. The analysis of our simulations, which extends the equal-mass binaries considered in Paper~I (i.e., Ref.~\cite{PhysRevD.96.104058}),
confirms many of the findings obtained in that previous study.

The fate of these binary mergers is either a nonrotating BS or a Kerr BH, as confirmed not only by global quantities and by the structure of the solution, but also by the gravitational QNMs of the remnant.
As in Paper~I, we once again find no evidence  that any of these binaries form a rotating BS or a scalar cloud
synchronized in its rotation about a spinning BH.
The asymmetry introduced by the unequal mass of the constituent stars perhaps makes the
formation of either of these remnants less likely.An analysis of the parameter space indicates the need to refine the initial configurations to assess whether a rotating remnant can be formed.

For a certain range of the initial angular momentum, the remnant undergoes a process similar to a tidal disruption in NSs, and a blob of scalar field is ejected. This process has already been observed in the equal-mass binaries of Paper~I, although the symmetry in that case induced the ejection of two blobs in opposite directions instead of a single blob observed here.
The ejection of a single blob produces a large recoil of the remnant. In our C012-C018 case,
the estimate of the recoil velocity is more than $10^4$ km/s, larger than the superkicks of
binary BHs and large enough to have significant implications for the expected dynamics of BSs
in the universe.
Because recent studies suggest that rotating solitonic BSs should be stable against the nonaxisymmetric instability~\cite{2021PhRvD.103d4022S}, the ejection of the scalar blob is not likely a result of such an instability.

We also evolve an unequal mass binary with one of the stars transformed to an anti-star. This
anti-star completely annihilates upon contact, dispersing scalar field to infinity.
The remaining scalar field settles to a lower mass, static BS in a clear demonstration of
the stability of these solutions under a strong perturbation.
	
Regarding the GWs emitted during the coalescence, we have found results comparable to those of binary BHs:  the $l=m=2$ mode of the strain is always dominant, although higher-order modes become more relevant as the mass ratio increases. We have estimated  how the ratio of the modes depends on the mass ratio during the last few orbits of the coalescence.

We have also analyzed the prospect of detecting differences between binary BS and binary BH gravitational signals
with ground interferometers. We have found that while the
merger portion of the signal is significantly different between the two classes of sources (at least if the final merger remnant is a BS), distinguishing between the two might be difficult with second-generation detectors due to degeneracies between merger and inspiral parameters. However, this task will ease considerably with third-generation interferometers, such as Cosmic Explorer or the Einstein Telescope.

Many interesting questions  remain to be addressed, especially regarding the final state of the remnant. Evolutions of solitonic BSs have yet to produce either a spinning BS or a synchronized scalar cloud. More accurate and longer simulations together with improved initial data may shed light on such questions, or perhaps some a priori analysis will indicate whether and under what conditions such end states will result.

\subsection*{Acknowledgments} 
It is a pleasure to thank Nico Sanchis-Gual for helpful comments and discussions on the manuscript, as well as Guillermo Lara for useful comments.
M.B, M.B., and E.B. acknowledge support from the European Union’s H2020 ERC Consolidator Grant “GRavity from Astrophysical to Microscopic Scales” (Grant No. GRAMS815673).
This work was supported by the EU
Horizon 2020 Research and Innovation Programme under the
Marie Sklodowska-Curie Grant Agreement No. 101007855.
This work was supported by the NSF under grants PHY-1912769 and PHY-2011383 (SLL).
P.P. acknowledges financial support provided under the European Union's H2020 ERC, Starting Grant agreement no.~DarkGRA--757480. We also acknowledge support under the MIUR PRIN and FARE programmes (GW-NEXT, CUP:~B84I20000100001), and from the Amaldi Research Center funded by the MIUR program ``Dipartimento di Eccellenza'' (CUP:~B81I18001170001).
This work was supported by European Union FEDER funds, the Ministry of Science, Innovation and Universities and the Spanish Agencia Estatal de Investigación grant PID2019-110301GB-I00 (C.P.).
Numerical calculations have been made possible through a CINECA-INFN agreement, providing access to resources on MARCONI at CINECA, as well as resources provided by XSEDE.
We acknowledge the use of CINECA HPC resources thanks to the agreement between SISSA and CINECA.
\clearpage	
\bibliographystyle{utphys}
\bibliography{biblio}

\end{document}